\documentclass{article}

\usepackage{PRIMEarxiv}

\usepackage[utf8]{inputenc} 
\usepackage[T1]{fontenc}    
\usepackage{url}            
\usepackage{booktabs}       
\usepackage{amsfonts}       
\usepackage{nicefrac}       
\usepackage{microtype}      
\usepackage{lipsum}
\usepackage{fancyhdr}       
\usepackage{graphicx}       

\usepackage{upgreek}
\usepackage{multicol,multirow}
\usepackage{amsmath,amssymb,amsfonts}
\usepackage{mathrsfs}
\usepackage{amsthm}
\usepackage[figuresright]{rotating}
\usepackage{appendix}
\usepackage[authoryear]{natbib}
\usepackage{ifpdf}
\usepackage[T1]{fontenc}
\usepackage{newtxtext}
\usepackage{newtxmath}
\usepackage{textcomp}
\usepackage{xcolor}
\usepackage{float}
\usepackage[colorlinks,allcolors=blue]{hyperref}
\definecolor{jourcolor}{cmyk}{1,0.57,0.01,0.38}
\hypersetup{
    colorlinks,%
    citecolor=jourcolor,%
    filecolor=jourcolor,%
    linkcolor=jourcolor,%
    urlcolor=jourcolor
}

\theoremstyle{definition}

\graphicspath{{media/}}     

\title{Numerical investigation of   mode failures in submerged granular columns}

\author{
  E. P. Montellà, J. Chauchat, C. Bonamy \\
  University of Grenoble Alpes, LEGI, G-INP, CNRS,
  38000 Grenoble, France\\
  \texttt{} 
  \\
  \And
  D. Weij \\
  Plaxis B.V., Computerlaan 14, 2628 XK Delft, The Netherlands\\
  \texttt{} \\
  \And
  G. H. Keetels \\
  Delft University of Technology, Mekelweg 2, Delft, The Netherlands\\
  \texttt{} \\
  \And
  T. J.  Hsu \\
  Center for Applied Coastal Research, University of Delaware, Newark, DE 19711, USA\\
  \texttt{} \\
}

\begin{document}
\maketitle

\textbf{\mathversion{bold}Abstract}In submerged sandy slopes, soil is frequently eroded as a combination of two main mechanisms: breaching, which refers to the retrogressive failure of a steep slope forming a turbidity current, and, instantaneous sliding wedges, known as shear failure, that also contribute to shape the morphology of the soil deposit.  Although there are several modes of failures, in this paper we investigate breaching and shear failures of granular columns using the two-fluid approach.  The numerical model is first applied to simulate small scale granular column collapses \citep{rondon2011granular}  with  different initial volume fractions to study the role of the initial conditions on the main flow dynamics. For loosely packed granular columns, the porous medium initially contracts and the resulting positive pore pressure leads to a rapid collapse. Whereas in initially dense-packing columns, the porous medium  dilates and negative pore pressure is generated stabilizing  the granular column, which results in a slow collapse. The proposed numerical approach  shows good agreement with the experimental data in terms of morphology and excess of pore pressure. Numerical results are extended to a large-scale application \citep{weij2020modelling,alhaddad2023stabilizing}  known as the breaching process. This phenomenon may occur naturally at coasts or on dykes and levees in rivers but it can also be triggered by humans during dredging operations. The results indicate that the two-phase flow model correctly predicts the dilative behavior and, the subsequent turbidity currents, associated to the breaching process.

\textbf{\mathversion{bold}Impact Statement}
Flow slides and breaching failures represent a major risk for buildings, roads, and other infrastructures. Additionally, in the dredging industry, the breaching process is used to extract sand and it may become unstable which may result in a loss of land. Although these phenomena are broadly observed in nature, the lack of understanding of the underlying physical processes is partly due to the absence of accurate experimental data and partly due to the complexity of the models. Laboratory experiments are challenging due to the impossibility of seeing through a dense suspension of sand particles. Numerical models, thus, may be a promising alternative to gain insight into these complex failure mechanisms and help engineers to better assess the risk of flow slides and
breaching. This contribution is one of the first attempts to develop a physically consistent two-phase flow model to predict the dynamics of a wide range of failure modes.

\section{Introduction}

Although granular flows are ubiquitous in nature and industrial applications, researchers still struggle to  completely understand 
the underlying physics of such flows. Difficulties arise when the granular material interacts with a viscous fluid:  the deformation of the soil skeleton induces changes on the fluid pressure field which, subsequently, affects the topology of the soil skeleton. Indeed, the complex inner nature of such flows is the main reason for the absence of simplified models  that describe immersed granular flows.

In this work, we explore the collapse of granular columns immersed in a viscous fluid. Special attention is given to \cite{rondon2011granular}, an experimental investigation of the role of the initial solid volume fraction on the dynamics of the granular collapse. \cite{rondon2011granular} observed that in initially loose packings, the excess of pore pressure built up and
enhanced a rapid mobility of the granular column, whereas in initially dense granular packings, negative  pore pressure developed within the porous medium increasing the shear resistance and, therefore, delaying the granular collapse.  This mechanism, due to the contracting/dilating behavior of granular material, is commonly known as pore pressure feedback and was first reported by \cite{iverson1997debris,iverson2005regulation} and largely studied experimentally \citep{iverson2000acute,pailha2008initiation,rondon2011granular,bougouin2018granular} and numerically \citep{bouchut2016two,wang2017dilatancy,lee2021two,montella2021two} using multiple configurations where the dense/loose granular material is sheared. Overall, on the collapse of immersed granular columns, literature \citep{rondon2011granular,lee2021two}  distinguishes two main processes largely linked to the initial volume fraction: on the one hand, initially loose packings lead to  shear failure forming a sliding wedge. On the other hand, dense granular packings are more prone to collapse through the breaching mechanism, that is the  process of front particles progressively released  as a turbidity current while the fluid penetrates the granular column to enhance the dilation of the medium.
Breaching failure carries potential danger in   submerge dense sandy soils. Indeed, \cite{beinssen2014field}  have  reported multiple large-scale failures due to the breach face slowly receding from the original position. Even though this paper focus on the breaching and shear plane failures, it is worth mentioning that other failure mechanisms, such as soil liquefaction, are also influenced by dilatancy effects \citep{youd1973liquefaction,prevost1985simple}. Liquefaction occurs in loose soils where a rapid particle rearrangement leads to a pore pressure buildup vanishing the effective stress.  Structures on liquefiable soils may have terrible consequences under earthquakes or other shear induced situations \citep{koutsourelakis2002risk}.

Several numerical studies on immersed granular collapses have been reported in the literature. \cite{kumar2017mechanics} explored the  effect of initial volume fraction on the dynamics of 2D granular collapses by means of  the Discrete Element
Method (DEM) coupled with the Lattice Boltzmann Method (LBM).
\cite{izard2018numerical} was able to reproduce three-dimensional granular collapses  with an immersed boundary method coupled with DEM. Similarly, \cite{xu2019analysis} and \cite{yang2019comprehensive} relied on smoothed particle hydrodynamics (SPH)-DEM and LBM-DEM approaches, respectively,  to study the process of submerged granular collapse. Most of previous work is computationally expensive and time consuming. Thus, affordable simulations are typically restricted to a low number of particles, which limits the range of applications, especially if one is  interested in large-scale applications such as coastal breaching or landslides. Alternatively, continuum approaches are more suited for large-scale problems but their accuracy highly depends on the  closure models and an adequate coupling between the fluid and the solid phase. On the one hand, the mixture model proposed by \cite{savage2014modeling} was capable of predicting the initial dynamics of the loose granular collapse but less satisfactory results were found for the dense granular collapse. This approach neglected the excess of pore pressure despite the fact that  pore pressure feedback mechanism plays a key role in the collapse. On the other hand, \cite{bouchut2017two} proposed a depth-averaged approach to model the dilatancy effects and pore pressure feedback mechanism of different submerged granular collapses. Based on the Eulerian-Eulerian framework, the first attempts by \cite{lee2015three, lee2018two} to model the intricate dynamics of granular column collapses obtained promising results. However, such models adopted simple elastic relationships to determine the solid pressure  ignoring the shear-induced volume changes which led to an insufficient description of the pore pressure feedback. \cite{shi2021theoretical} and \cite{lee2021two} went one step further and proposed modifications of the elastic equation  that captured the sheared-induced volume changes and reproduced the granular collapse dynamics with great success. Previous studies are mainly focused on the morphology of the deposit and the pore pressure feedback mechanism of small-scale granular collapses. Yet, experimental studies \citep{van1999breaching,van2002importance,eke2011field} show that the breaching process forming turbidity currents is a crucial mechanism to reproduce practical applications such as dredging engineering  or protection measures against coastal erosion \citep{van2002importance,beinssen2014field,shipway2015risks,mastbergen2019watching}.  Accordingly, it seems reasonable to upscale the problem and examine whether the collapse is driven by the same dynamics observed in Rondon's experiments \citep{rondon2011granular} or, conversely, other physics apply. Although the studies on breaching are scarce \citep{van1999breaching,you2012dynamics,you2014mechanics,alhaddad2020large,weij2020modelling,alhaddad2023stabilizing},  most of the laboratory experiments reported that the granular columns manifest a dual mode slope failure. Instead of observing a progressive erosion induced by the turbidity currents of the breaching process, the evolution of the granular column is governed by a combination of breaching and occasional sliding wedges. 
\cite{you2014mechanics} suggests that slide failures take place when the negative pore pressure developed in the porous medium is not enough to keep the shear resistance  against slide failure.  \cite{you2014mechanics} remarked that slides are associated to a drop in excess of pore pressure, and, the magnitude of the jump is related  to the size of the slide. Additionally,   experiments  \citep{ lee2021onset,lee2022multiphase} have reported that dense granular packings exhibit shear failure for coarse particles and breaching for fine particle. It is worth noting that experiments investigating the breaching process are limited in size because large-scale experiments are not affordable. Thus, numerical simulations arise as a potential alternative  to study the physics of the breaching process in large-scale  applications. \cite{breusers1977local} introduced the concept of active wall velocity defined as the horizontal velocity at which steep slopes move  due to the breaching process. Since then,  several breaching erosion models have been proposed based on different closures for the  wall velocity. A quasistatic 1D depth-averaged approach developed by \cite{mastbergen2003breaching} was used to investigate the breaching process showing that turbidity currents  can be strong enough to periodically flush large deposits of sediments from canyons.  \cite{eke2011field} considered a similar model to study another flushing event in submarine canyons. More recently, \cite{van2015slope} proposed a 2D drift-flux model  based on the Reynolds averaged Navier–Stokes equations  to study the stability of the breaching process.  Previous studies, nonetheless, are  subjected to some limitations. Firstly,
current models do not account for slide failures, and, secondly, they neglect the evolution of soil properties. Therefore, in order to further investigate the breaching process,  numerical models should not only be able to capture the turbidity currents, but also the transition from static to yielding soil and the effects of the pore pressure feedback.

In this work, we first validate the elastoplastic model presented in \cite{montella2021two} using the Rondon's experiments \citep{rondon2011granular} as reference. Then, we report on the application of the numerical model to study the effects of the breaching process of a granular column collapse $\sim$35 times larger \citep{weij2020modelling} than the one from Rondon's experiments.

\section{Mathematical formulation}\label{methods}

The governing equations for the Eulerian-Eulerian two-phase formulation are shown below alongside with the closure forms for drag force, turbulence model and stresses for the fluid and particle phases.

\subsection{Two-phase flow governing equations}\label{intro_eq}

The mass continuity equations for the solid  and fluid phase are written as follows:
\begin{equation}\label{eq:mass_eq}
 \frac{\partial \phi }{\partial t} + \nabla  \cdot ( \mathbf{u^s} \phi )= 0,
\end{equation}

\begin{equation}\label{eq:mass_fluid_eq}
 \frac{\partial (1-\phi) }{\partial t} + \nabla  \cdot ( \mathbf{u^f} (1-\phi) )= 0
\end{equation}

Here, $\phi$, $ \mathbf{u^s}$ and $ \mathbf{u^f}$ are the solid volume faction,  the particle phase velocity and   the fluid  phase velocity, respectively.

The momentum conservation equations for the solid phase and fluid phase are written as: 
\begin{equation}\label{eq:mom_solid}
\begin{split}
\rho^s \phi  \left[ \frac{\partial \mathbf{u^s} }{\partial t} +  \mathbf{\nabla} \cdot  \left(\mathbf{u^s}\otimes  \mathbf{u^s} \right)\right]= \phi(\rho^s-\rho^f) \mathbf{g} + \dfrac{(1-\phi) \rho^f \nu^f}{K} (\mathbf{u^f} - \mathbf{u^s})   - \nabla p^s +   \nabla \cdot  \boldsymbol{\tau^s}  - \phi \nabla p^f,
\end{split}
\end{equation}

\begin{equation}\label{eq:mom_fluid}
\begin{split}
\rho^f (1-\phi)  \left[\frac{\partial \mathbf{u^f} }{\partial t}+  \mathbf{\nabla} \cdot \left( \mathbf{u^f}\otimes  \mathbf{u^f} \right)\right]=  \dfrac{(1-\phi) \rho^f \nu^f}{K} (\mathbf{u^s} - \mathbf{u^f})   +  \nabla \cdot  \boldsymbol{\tau^f}   - (1-\phi) \nabla p^f,
\end{split}
\end{equation}
where $\rho^s$ is the solid density, $\rho^f$ is the fluid density, $\nu^f$ stands for the fluid kinematic viscosity, $\otimes$ is the outer product of two vectors, $p^f$ is the excess  of pore  pressure defined as the difference between the pore pressure and the hydrostatic pressure,  $p^s$ is the solid pressure, $\boldsymbol{\tau^s} $ is the granular shear stress, $\boldsymbol{\tau^f} $ is the fluid  shear stress and $K$ is the permeability of the porous medium.

This work uses the approach of  \cite{engelund1953laminar} to compute the permeability based on the pressure drop in steady porous flow:

\begin{equation}\label{eq:dragEngelund}
K=  \frac{d^2  (1-\phi)}{\alpha_E \phi^2 } +\frac{d \nu^f (1-\phi)^2}{\beta_E   \left\| \mathbf{u^f} - \mathbf{u^s} \right\| },
\end{equation}

where  $d$ is the mean particle diameter. According to \cite{burcharth1991stationary,burcharth1995one}, $\alpha_E$ ranges from 780 (for uniform and spherical particles) to 1500 or more (for irregular and angular grains), while  $\beta_E$
ranges from 1.8 (for uniform and spherical particles) to 3.6 or more (for irregular and angular grains). The choice of  \cite{engelund1953laminar} model over other approaches employed for high packing fractions (such as \cite{ergun1952fluid}) lies on  the advantage to adjust the factors $\alpha_E$  and $\beta_E$ to accurately represent the permeability of soils made up of irregular shaped particles.

\subsubsection{Solid and fluid stress}\label{stresses}

The fluid phase shear stress is expressed as the sum of the Reynolds stresses due to  turbulent fluctuations ($\boldsymbol{R^{t}}$) and the viscous stresses ($\boldsymbol{r^f}$):

\begin{equation}\label{eq:fluid_shear_stress}
\boldsymbol{\tau^f}= \boldsymbol{R^{t}}+\boldsymbol{r^{f}},
\end{equation}

where the  Reynolds stress tensor $\boldsymbol{R^{t}}$ is modeled as:

\begin{equation}\label{eq:fluid_shear_stressReynolds}
\boldsymbol{R^{t}} = 2 \rho^f (1
-\phi) \left [  \nu^{t} \boldsymbol{S^f} -\frac{1}{3}k  \boldsymbol{I} \right ],
\end{equation}

and the viscous stress is written as follows:

\begin{equation}\label{eq:fluid_shear_stressVisc}
\boldsymbol{r^{f}}= 2 \rho^f (1
-\phi)    \nu^{mix} \boldsymbol{S^f},
\end{equation}

where 

\begin{equation}\label{eq:deviatoric_fluid}
\boldsymbol{S^f}= \frac{1}{2} \left ( \boldsymbol{\nabla} \mathbf{u^f}  + (\boldsymbol{\nabla} \mathbf{u^f})^T  \right ) - \frac{1}{3} tr(\boldsymbol{\nabla}  \mathbf{u^f}),
\end{equation}

is the deviatoric and symmetric part of the velocity gradient for the fluid phase, $k$ is the turbulent kinetic energy,  $\nu^{t}$ stands for the eddy viscosity calculated with a  turbulence closure model (see section \ref{turubulence}) and  $\nu^{mix}$ is the mixture viscosity, which according to \cite{boyer2011unifying}, can be computed with the following   phenomenological expression:

\begin{equation}\label{eq:dynamicViscosity}
\nu^{mix}=\nu^{f}  \left [1 + 2.5 \phi \left ( 1- \frac{\phi}{\phi_{max}}\right )^{-1} \right ],
\end{equation}

where $\phi_{max}$ is the maximum solid volume fraction set to  $\phi_{max}=0.625$ for spheres.

Following \cite{cheng2017sedfoam,chauchat2017sedfoam}, the  solid phase pressure $p^s$ is defined as the sum of a viscous shear-rate dependent pressure $p_s^s$ and the contribution of enduring contacts $p_c^s$:

\begin{equation}\label{eq:particle_pressure}
p^s=p^s_s+p^s_c,
\end{equation}

$p_c^s$  is proportional to the difference  between the  solid volume fraction $\phi$ and the reference solid fraction $\phi_{pl}$ where dilatancy effects are embedded:

\begin{equation}\label{eq:J_and_J2}p^s_{c}=
  \begin{cases}
    \! 
    \begin{alignedat}{2}
      &0 \quad \quad \quad \quad  \quad \quad \quad \quad \quad \quad     \phi  < \phi_{pl}
      \\
      &E \frac{(\phi-\phi_{pl})^3}{(\phi_{rcp}-\phi)^5} \quad  \quad  \quad \quad\phi  \geq \phi_{pl},
    \end{alignedat}
  \end{cases}
\end{equation}

where $\phi_{rcp}$ is the random close packing volume fraction.  We adopt the value for sphere packings ($\phi_{rcp}=0.625$). It's important to note that Eq. \ref{eq:J_and_J2} is based on the work of \cite{johnson1987frictional}, which assumes a constant value called the random loose packing fraction ($\phi_{rlp}$ instead of $\phi_{pl}$). However, to accurately account for the effects of dilatancy, we need to consider initial and transient packing fractions that are different from the random loose packing fraction. The initial volume fraction ($\phi_{o}$) is calculated as the average volume fraction throughout the height of the bed, which is given by:

\begin{equation}\label{eq:average_initial}
\phi_{o}=\frac{1}{h_o}\int_{o}^{h_o} \phi(y,t=0) ,dy 
\end{equation}

Here, $h_o$ represents the lowest position above which $\phi\le\phi_{top}=0.53$. By adjusting the initial plastic volume fraction ($\phi_{pl,t=0}$), which remains constant during the process of gravitational deposition for preparing the sample, we can achieve different initial volume fractions. Higher values (in this study, $\phi_{pl,t=0} = 0.609$ to match \cite{rondon2011granular}) result in initially dense granular beds ($\phi_o \approx 0.61$), while lower values (in this study, $\phi_{pl,t=0} = 0.54$ to reproduce \cite{rondon2011granular}) yield initially loose granular packings ($\phi_o \approx 0.55$).

Once the system reaches an equilibrium state and the numerical sedimentation is complete,  the granular collapse begins and $\phi_{pl}$ evolves to account for the plastic effects. Following \cite{montella2021two}, the plastic effects that arise from local rearrangements during shearing deformations are captured as an increment/reduction of $\phi_{pl}$, which further changes the solid pressure (see Eq. \ref{eq:J_and_J2}). More specifically, the expression that governs the evolution of $\phi_{pl}$   is:

\begin{equation}\label{eq:volum_strain_temporal}
\frac{\partial\phi_{pl}}{\partial t} + \mathbf{u^s} \cdot \nabla \phi_{pl}  = - \phi_{pl} \delta  \vert\vert \boldsymbol{S^s}\vert\vert, 
\end{equation}

where  $\boldsymbol{S^s}$ is the deviatoric strain rate of the solid phase computed as in Eq.\ref{eq:deviatoric_fluid} but replacing the fluid velocity with the solid velocity  and $\delta$ is the dilatancy coefficient defined as:

\begin{equation}\label{eq:dilatancy}
\delta = K_1 (\phi-\phi_{\infty}), 
\end{equation}

where $K_1$ is a calibration parameter and $\phi_{\infty}$ stands for  the equilibrium volume fraction. As reported by \cite{roux1998texture,pailha2009two}, the linear variation of the dilatancy angle with the volume fraction as written in Eq. \ref{eq:dilatancy} is derived from the linerization of  the dilation rate:

\begin{equation}\label{eq:dilatancyDef}
\nabla \cdot \mathbf{u^s} = \frac{1}{\phi} \frac{d \phi}{dt} = \delta \vert\vert \boldsymbol{S^s}\vert\vert 
\end{equation}

$\phi_{\infty}$ is modeled as a function of the particle pressure  and  the shear rate through the viscous number: 

\begin{equation}\label{eq:phi_viscous} 
\phi_{\infty} = \frac{\phi_{c}}{1+ I_v^{1/2}},
\end{equation}

 where   the viscous number  ($I_v$) is defined as:
\begin{equation}\label{eq:Iv}
I_v =  \dfrac{ \rho^f \nu^f \vert\vert \boldsymbol{S^s}\vert\vert }{p^s},
\end{equation}

and $\phi_{c}$ is the critical volume fraction in quastistatic shear ($I_v \rightarrow 0$).

In this work  $ \delta $  is limited to a range of $ -0.4 \leq \delta \leq 0.4$ which falls into physical values proposed by previous works \citep{pouliquen1996onset,iverson2014depth,alshibli2018influence}.  The influence of the dilatancy prefactor $K_1$ will be studied in section \ref{sensitivity}.

Eq. \ref{eq:phi_viscous} suggests that different values of the equilibrium solid volume fraction are expected in transient conditions, such as the onset of the granular collapse, the fully developed flow and the final arrest. In the context of a granular column collapse, steady flows are unlikely: viscous forces are expected to slow down the granular flow before it is fully developed, therefore, dilatancy effects will arise provided that $\phi \neq \phi_\infty$ and $\vert\vert \boldsymbol{S^s}\vert\vert  > 0$. It is noteworthy that despite the fact the value of $\phi_{pl}$ is just a numerical parameter, the consequences of the changes in $\phi_{pl}$ are completely physical.  Indeed, this model not only extends the critical state soil mechanics to a rate-dependent critical state ($\phi_{\infty}$) but also leads to an increase or a decrease of the granular pressure depending on the initial packing and, consequently, to pressure-driven expansion or compaction of the solid phase under shear conditions. Furthermore, the value of $\phi_{pl}$  remains unbounded  because   during the granular flow  it cannot decrease or increase indefinitely. This is because for loose materials (low $\phi_{pl}$ value), the dilatancy coefficient is negative, and according to Eq. \ref{eq:volum_strain_temporal}, $\phi_{pl}$ must increase. Similarly, for dense packings, it is also not possible for $\phi_{pl}$ (initially large) to continuously increase because positive dilatancy coefficients lead to a reduction in $\phi_{pl}$.

The expression for the shear rate-dependent pressure induced by collisional  interactions was derived by \cite{chauchat2017sedfoam} inverting Eq. \ref{eq:phi_viscous} to give the rate-dependent normal stress $p^s_{\infty}$:

\begin{equation}\label{eq:ps_viscous}
p^s_{\infty} = \rho^f \nu^f \vert\vert \boldsymbol{S^s}\vert\vert \left ( 1- \frac{\phi_{c}}{\phi} \right )^{-2}
\end{equation}

However, as suggested by \cite{montella2021two}, $p^s_{\infty}$ is consistently defined  to be the stationary shear-induced pressure whereas the actual pressure is supposed to converge asymptotically to that value with accumulated strain, therefore, the following equation governs the progressive mobilization of $p_s^s$:

\begin{equation}\label{eq:pss_relax}
\frac{\partial p^s_s}{\partial t} + \mathbf{u^s} \cdot \nabla  p^s_s  = -   K_2 (  p^s_s-p^s_{\infty}) \vert\vert \boldsymbol{S^s}\vert\vert
\end{equation}

In short, dilatancy is a complex physical phenomenon and our approach consists in decomposing  its effect into the enduring contact pressure $p_c^s$ and the shear-rate dependant pressure $p_s^s$. The former pressure ($p_c^s$) is closely related to the microstructure. For instance, in initially dense packings, the particles are interlocking and are not able to move freely. Therefore, the grains need to be rearranged to allow shearing deformations. During the grain reorganization, the contacts become stronger which comes with an increment of the  enduring contact pressure $p_c^s$. The shear-rate dependant pressure $p_s^s$, on the contrary,  is derived from the $\mu(I_v)$ rheology. \cite{boyer2011unifying} showed the granular medium dilates when increasing the shear rate ($I_v \uparrow$) which is accompanied by an increment of the solid pressure that scales with the fluid viscosity and the shear-rate.

\cite{jop2006constitutive} proposed that the ratio between shear stress and pressure can be scaled by
the inertial number, $I=\frac{d \vert\vert \boldsymbol{S^s}\vert\vert  }{\sqrt{p^s / \rho^s}}$, defined as the ratio between the macro and micro time scale of granular flow. \cite{cassar2005submarine} followed a similar approach using the viscous number $I_v$ to model granular flows immersed in viscous fluids. Even though the present work utilizes the viscous number $I_v$ because the analyzed granular collapses are found to be in the viscous regime (the Stokes number $St=\frac{\rho^s d^2 \vert\vert \boldsymbol{S^s}\vert\vert }{\rho^f \nu^f}$ ranges from $St=0.005$ to $St=0.3$ depending on the granular collapse), different scenarios or real large-scale events, where the rheology belongs to an inertial regime rather than a viscous regime, may be studied with the present numerical model simply adapting the constitutive laws for the inertial number $I$ as reported by   \cite{montella2021two}.

In this work, the solid shear stress $\boldsymbol{\tau^s}$ is   proportional to the solid pressure following a frictional law depending on the viscous number $I_v$:
\begin{equation}\label{eq:particle_shear_stress_coulomb}
\boldsymbol{\tau^s}= \mu(I_v) p^s \frac{\boldsymbol{S^s}}{ \vert\vert \boldsymbol{S^s}\vert\vert},
\end{equation}

where $\mu(I_v)$ is the friction coefficient for a certain shear state described in \cite{boyer2011unifying} as:

\begin{equation}\label{eq:muIVEq}
\mu(I_v)=  \mu_s + \dfrac{ \Delta \mu }{ I_o/I_v + 1},
\end{equation}

where the empirical material constants correspond to  the static friction coefficient $\mu_s$, the dynamic friction coefficient $\Delta \mu$, and the reference viscous number $I_o$.

In order to have an expression for  $\tau^s$ resembling the definition for the fluid shear stress, the shear stress due to frictional contacts can be rewritten as: 

\begin{equation}\label{eq:particle_shear_stress}
\boldsymbol{\tau^s}= \rho^s \nu^s \boldsymbol{S^s},
\end{equation}

where $\nu^s$ is the frictional shear viscosity:

\begin{equation}\label{eq:frictional_viscosity_regularized}
\nu^s=  \frac{ \mu(I_v)  p^s}{ \rho^s \left ( \vert\vert \boldsymbol{S^s}\vert\vert^2+\lambda^2_r \right )  ^{1/2}},
\end{equation}
where $\lambda_r$ is a regularization parameter from \cite{chauchat2014three} taken equal to $\lambda_r=10^{-6}$ s$^{-1}$. Moreover, $\nu^s$ is limited to be smaller than  $\nu^s_{max}=10^{5}  m^2 s^{-1}$ to avoid numerical issues.

\subsubsection{Turbulence model}\label{turubulence}

To model the turbulent eddy viscosity $\nu^t$, the $k-\epsilon$ model \citep{hsu2004toward,cheng2017sedfoam} is used in this study. $\nu^t$ is computed as:

\begin{equation}\label{eq:turbVisc}
\nu^t = C_{\mu} \frac{k^2}{\epsilon},
\end{equation}

where $C_{\mu}=0.09$ is an empirical coefficient, $k$ is the turbulent kinetic energy and $\epsilon$ is the dissipation rate of turbulent kinetic energy.  The turbulent kinetic energy $k$ is determined with the following transport equation:

\begin{equation}\label{eq:TKE}
\frac{\partial k}{\partial t} +u^f_j  \frac{\partial k}{\partial x_j}  = \frac{R^{t}_{ij}}{\rho^f} \frac{\partial u^f_i}{ \partial x_j} + \frac{\partial}{\partial x_j} \left [  \left (\nu^f +  \frac{\nu^t}{\sigma_k} \right ) \frac{\partial k}{\partial x_j} \right ] -\epsilon - \frac{2 K (1-t_{mf}) \phi k }{\rho^f} ,
\end{equation}
where $R^t_{ij}$ is the Reynolds stress tensor, $\sigma_k$ is an empirical coefficient and $t_{mf}$ is a parameter that characterizes the degree of correlation between particles and fluid velocity fluctuations modeled as:  $t_{mf} = e^{-B \cdot St}$, in which $B$ is an empirical coefficient chosen as $B = 0.25$ and $St$ is the Stokes number defined as: $St=\frac{t_p}{t_l}$ where $t_p=\rho^s K \phi / ((1-\phi) \rho^f \nu^f)$ is the particle response  and $t_l=k/(6 \epsilon)$ is the characteristic time scale of energetic eddies. 

The balance equation for the dissipation rate of turbulent  kinetic energy is written as:

\begin{equation}\label{eq:epsilonEq}
\begin{split}
\frac{\partial \epsilon}{\partial t} +u^f_j  \frac{\partial \epsilon}{\partial x_j}  = C_{1 \epsilon} \frac{\epsilon}{k} \frac{R^{t}_{ij}}{\rho^f} \frac{\partial u^f_i}{ \partial x_j} + \frac{\partial}{\partial x_j} \left [  \left (\nu^f +  \frac{\nu^t}{\sigma_{\epsilon}} \right ) \frac{\partial \epsilon}{\partial x_j} \right ] -C_{2 \epsilon}\frac{\epsilon^2}{k} -C_{3 \epsilon}\frac{\epsilon}{k} \frac{2   \nu^f (1-t_{mf})  k }{K} 
\end{split}
\end{equation}

In this work, the values of the turbulent empirical coefficients are found in table \ref{tab:table_TurbulentCoef}.

\begin{table}
  \begin{center}
\def~{\hphantom{o}}
  \begin{tabular}{cccccccc}
      \textbf{Parameter} &   $\sigma_k$ &   $\sigma_{\epsilon}$   & $C_{1 \epsilon}$ & $C_{2 \epsilon}$ & $C_{3 \epsilon}$  & $C_{4 \epsilon}$    & $C_{\mu}$  \\[2pt]
  \textbf{Value} & 1 & 0.77 & 1.44  & 1.92 & 1.2 & 1 & 0.09\\
  \end{tabular}
  \caption{Empirical coefficients for the $k - \epsilon$ turbulence model taken  from  \cite{chauchat2017sedfoam}.}
  \label{tab:table_TurbulentCoef}
  \end{center}
\end{table}

\subsection{Numerical implementation}\label{implementation}

Simulations are conducted with the open-source software SedFoam, a two-phase flow solver  used for sediment transport applications \citep{chauchat2017sedfoam,mathieu2019two,montella2021two,chassagne2023frictional,tsai2022eulerian,mathieu2022numerical} based on the open-source finite volume library OpenFOAM \citep{jasak2020practical} (v2212 release from ESI). The solver is available for download on  \footnote{https://github.com/SedFoam/SedFoam}{\href{https://github.com/SedFoam/sedfoam}{GitHub}}.   Several interpolation/discretization techniques can be used to evaluate the face fluxes. Table \ref{tab:table_schemes}  shows schemes used for temporal and spatial discretization. It is worth noting that \textit{Gauss upwind} scheme is only adopted for the divergence  discretization of $\phi_{pl}$ and  $p^s_s$ fields. In order to resolve the velocity-pressure coupling, we rely on  the pressure-implicit split-operator (PISO) algorithm.

\begin{table}
  \begin{center}
  \begin{tabular}{cc}
\hline
\multicolumn{1}{c}{\textbf{Description}}                              & \multicolumn{1}{c }{\textbf{Scheme}}       \\ \hline
Time discretization                                                     & \textit{Euler}                          \\
Gradient term  discretization & \textit{Gauss linear}                      \\
Divergence operators                                                    & \textit{Gauss limitedLinear, Gauss upwind} \\
Laplacian operator                                                      & \textit{Gauss linear corrected}
\end{tabular}
  \caption{Numerical schemes for the interpolation of the convective fluxes.}
  \label{tab:table_schemes}
  \end{center}
\end{table}

\section{Results}

In this section, the two-phase flow model is used to reproduce initially loose and dense granular column collapses. The first part of this section consists of reproducing \cite{rondon2011granular}  granular collapses. Although, a broad number of works  \citep{wang2017dilatancy,izard2018numerical,jing2019flow,sun2020numerical,meng2021eulerian,riffard2022numerical,polania2022collapse} have predicted the main features of  granular column collapses immersed in a viscous fluid,  only a few \citep{wang2017two,yang2020pore} have successfully captured the dynamics of Rondon's experiments for both initially loose and dense granular columns, and even fewer, have done it with a continuum approach \citep{bouchut2017two,si2018development,baumgarten2019general,lee2021two,rauter2021compressible,shi2021theoretical,phan2022modelling}. The dilatancy model presented herein was able  to capture the pore pressure feedback in 1D and 2D granular avalanches \citep{montella2021two} with reasonably good agreement. The granular column collapse is, thus, used to extend the model to a more complex and realistic configuration. Additionally, a sensitivity analysis is summarized in this section to underline the parameters that govern the dynamics of the granular collapse complementing previous work on this topic \citep{lee2021two,rauter2021compressible}. Once the model is validated, the second part of this section will be devoted to gain insight into the breaching process. In order to achieve this goal, the laboratory experiments of \cite{weij2020modelling,alhaddad2023stabilizing} will be  reproduced numerically using a 2D approach.

\begin{figure}
\centering
\includegraphics[width=0.5\textwidth]{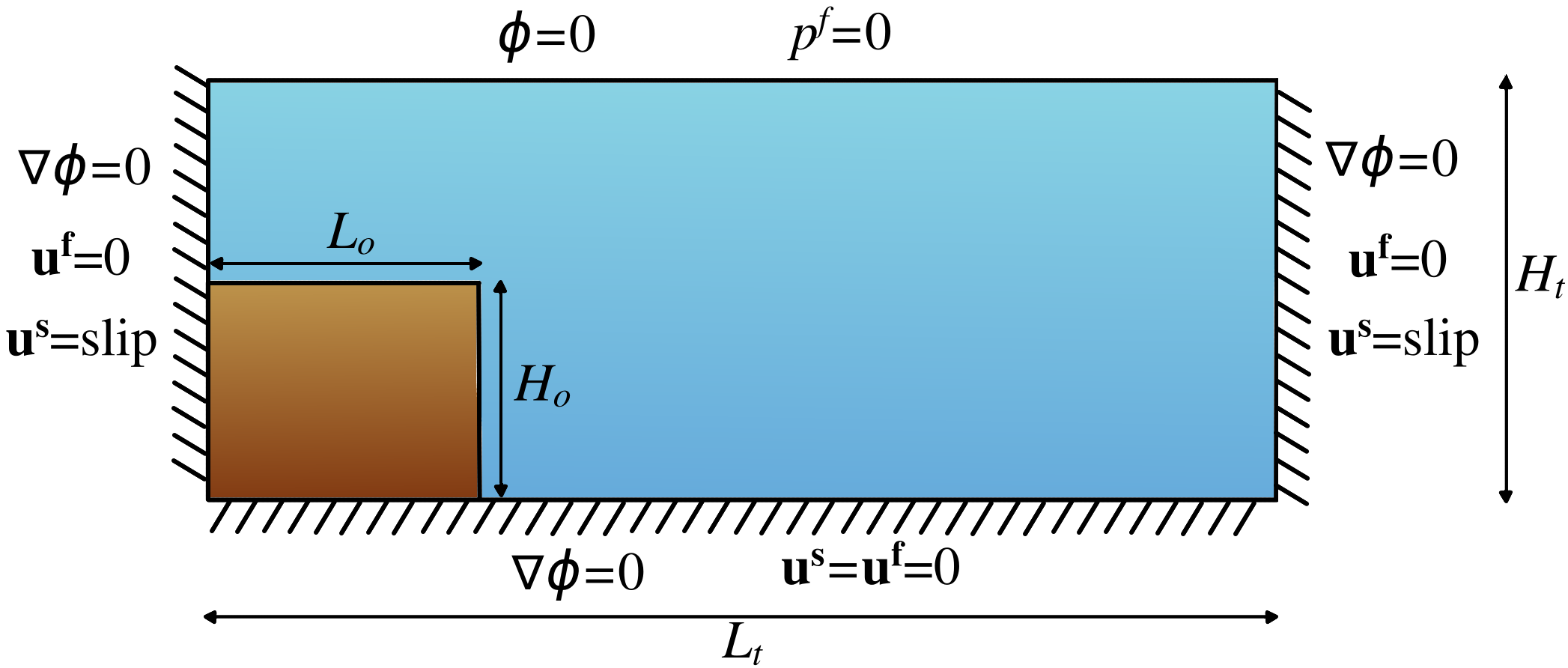}
\caption{\label{fig:SetUpRondon}Numerical setup
 to predict the granular collapse of \cite{rondon2011granular}.}
\end{figure}

\begin{table}
  \begin{center}
\def~{\hphantom{o}}
  \begin{tabular}{cccccc}
      \textbf{Parameter} &  \textbf{symbol} & \textbf{SI unit}   & \textbf{Value} \\[3pt]
        Solid density & $\rho^s$   & $kg/m^3$   & 2500 \\
       Fluid density & $\rho^f$  & $kg/m^3$   & 1010\\
       Fluid viscosity & $\nu^f$& $m^2/s$ & $1.2 \cdot 10^{-5} $\\
       Particle diameter & $d$ & $m$ & $225 \cdot 10^{-6} $ \\
       Height of the column & $H_o$ & $m$ &  0.042 dense , 0.048 loose \\
       Length of the column & $L_o$ & $m$ & 0.06\\
       Height of the tank & $H_t$ & $m$ &  0.08\\
       Length of the tank & $L_t$ & $m$ & 0.20 dense , 0.25 loose \\
  \end{tabular}
  \caption{Physical and geometrical variables used in the numerical simulations.}
  \label{tab:table_variables}
  \end{center}
\end{table}

\begin{table}
  \begin{center}
\def~{\hphantom{o}}
  \begin{tabular}{cccccccc}
      \textbf{Parameter} &   $\mu_s$ &  $\Delta \mu$   & $I_o$ & $K_1$ & $K_2$ & $\phi_c$  \\[2pt]
  \textbf{Value} & 0.425 & 0.34 & 0.004  & 40 & 1 for loose, 0.01 for dense & 0.57\\
  \end{tabular}
  \caption{Rheological and numerical parameters used to reproduce \cite{rondon2011granular}.}
  \label{tab:table_Num}
  \end{center}
\end{table}

\subsection{Collapse of a granular column - Rondon's experiments}\label{RondonSect}

The experiments performed by 
\cite{rondon2011granular} investigated  the collapse of a granular column in a viscous liquid.   Initially dense columns resulted in negative pore pressures that slowed down the collapse, while in loose granular packings, the collapsing process was triggered instantaneously with positive pore pressures that enhanced a rapid flow. Although several volume fractions and aspect ratios were analyzed, only two representative cases (an initially loose column with $\phi_o \approx 0.55$ and an initially dense packing with  $\phi_o \approx 0.61$) will be considered in this work.  The experimental setup consisted of  a tank with a length of $0.7 m$, and a width and height of $0.15 m$. A removable gate was placed vertically and glass beads were poured behind the gate.  The rest of the physical and geometrical parameters have been taken from  \cite{rondon2011granular} (see table \ref{tab:table_variables}). During the experiments, the
excess of pore pressure was measured at the bottom at $2cm$ from the left side of
the tank. The numerical setup is presented in figure \ref{fig:SetUpRondon}. The  numerical domain  is decomposed into square cells of $0.41mm \times 0.41mm $.

\begin{figure}
\centering
\includegraphics[width=0.8\textwidth]{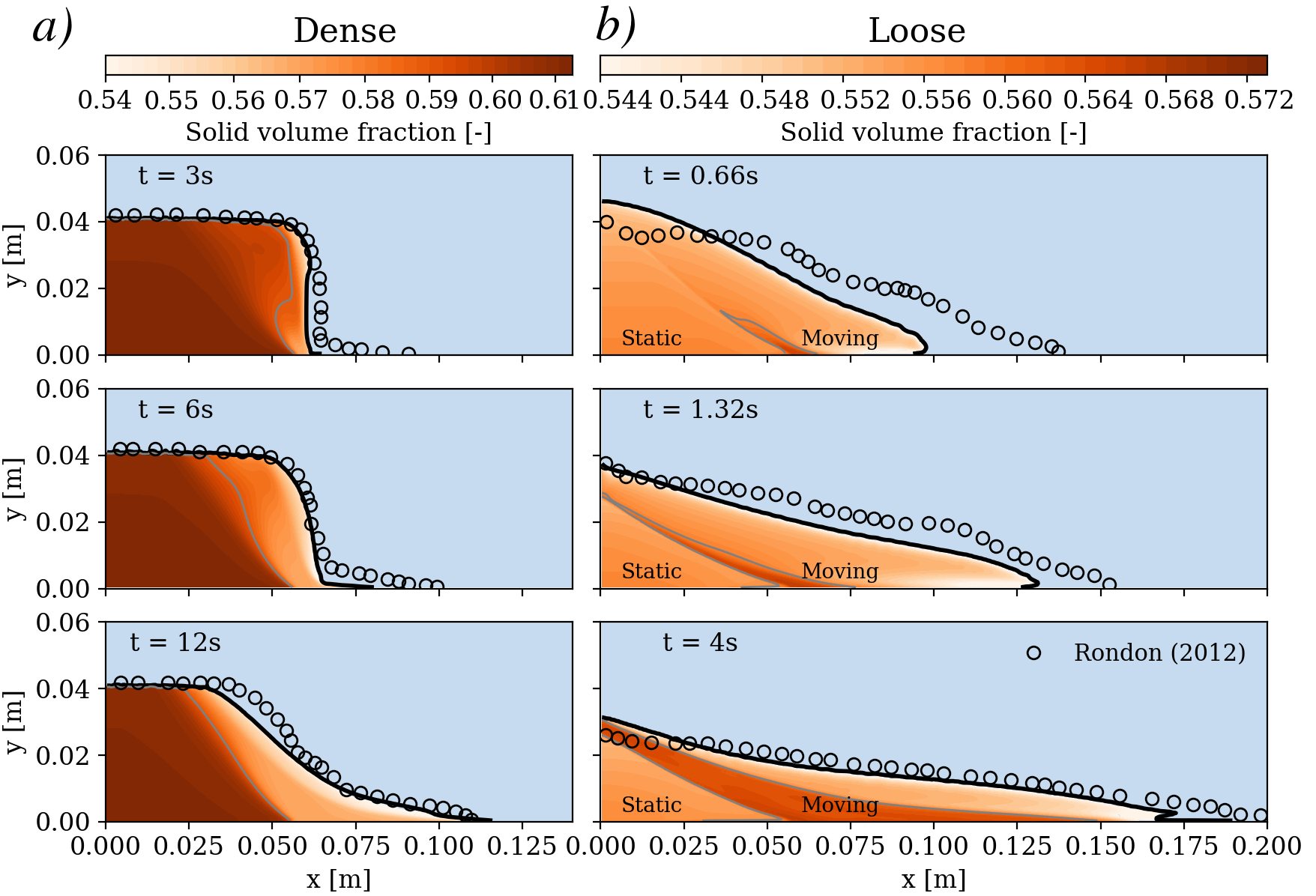}
\caption{\label{fig:Morpholohy}Evolution of the morphology and solid volume fraction during the collapse of an initially a) dense and b) loose column.  A gray line is included to illustrate the evolution of the isoline with the initial volume fraction ($\phi_o = 0.55$ for the initially loose column and $\phi_o = 0.61$  for the initially dense column).}
\end{figure}

According to \cite{rondon2011granular}, the internal friction angle is around $20^o$ and the critical volume fraction is $\phi_c=0.58$. In this work, however, we adopt the rheological parameters as in \cite{montella2021two}  with a   critical volume fraction of $\phi_c=0.57$. The rest of  rheological coefficients and calibration parameters are  summarized in table \ref{tab:table_Num}.   The permeability in Eq. \ref{eq:mom_fluid} is modeled according
to Engelund's model presented before with the following coefficients $\alpha_E=780$ and $\beta_E=1.8$ corresponding to smooth spherical particles. This set of parameters led to the best fit to the experimental results. The influence of these parameters is further discussed in section \ref{sensitivity}. Moreover, one may argue that $K_2$ values should be taken the same for both loose and dense scenarios. Ideally, $K_2$ should be $1$ so the relaxation time is dominated solely by the shear rate (see Eq. \ref{eq:pss_relax}). However, numerical instabilities in the dense case forced us to set an additional relaxation.  The choice of $K_2=0.01$, nonetheless, has a minor influence on the results because the inherent slow flow  of dense granular collapse is driven mainly by changes in the contact pressure (see section \ref{K2Sensi}). Finally, numerical simulations of Rondon's experiments belong to the viscous regime for the fluid phase, consequently, the turbulent viscosity is set to zero ($\nu^t=0$) for simplicity.

\subsubsection{Morphology}

Figure \ref{fig:Morpholohy} shows the evolution of the deposit shapes during the granular column collapses. As reported by \cite{rondon2011granular}, the dynamics of the granular column collapse are very different depending on the initial volume fraction: initial dense granular packings are mobilized slowly and show short run-out distances (see figure \ref{fig:Morpholohy}a). On the contrary, initially loose granular packings are characterized by a rapid flow and elongated fronts (see figure \ref{fig:Morpholohy}b).

\begin{figure}
\centering
\includegraphics[width=0.8\textwidth]{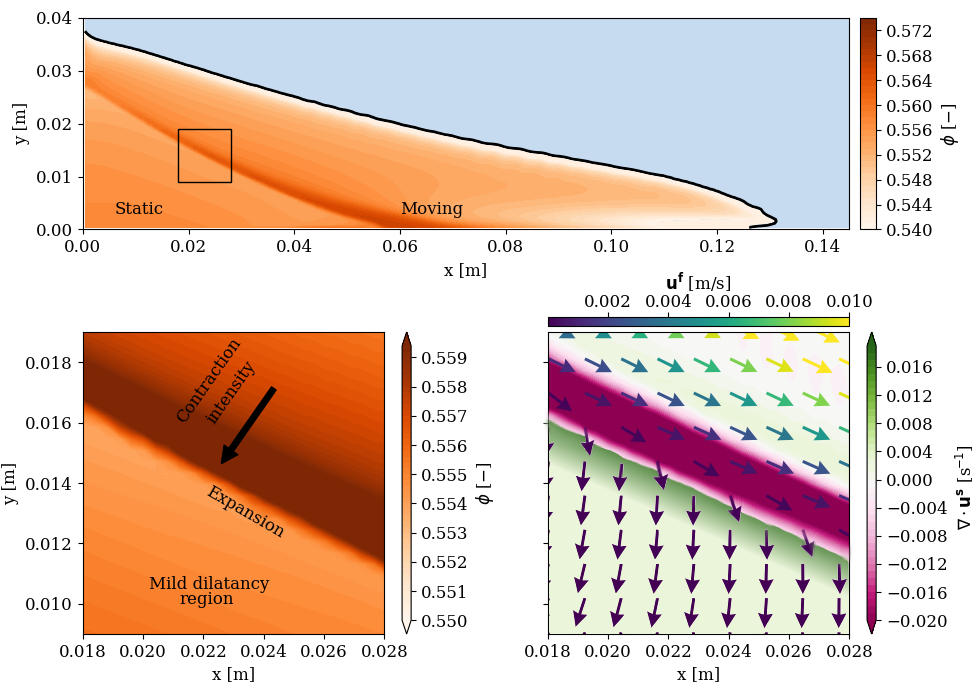}
\caption{\label{fig:GradAlphaLoose} a) Solid volume fraction and  zoom-in view along the failure surface with b) detailed volume fraction and c)  divergence of the solid phase velocity  and fluid flow field for the initially loose granular column. It must be noted that the arrows displayed in the subplot represent the magnitude of the fluid velocity by their color and not their size.}
\end{figure}

In addition to  the dynamics of the granular flow, dilatancy effects are also illustrated in figures   \ref{fig:Morpholohy} as changes in the solid volume fraction. The sheared region in figure  \ref{fig:Morpholohy}a is expanding progressively from $\phi_0 \approx 0.61$ to $\phi \approx 0.57$. Conversely, figure \ref{fig:Morpholohy}b shows a contraction of the sheared region from $\phi_0 \approx 0.55$ to $\phi \approx 0.565$. It is worth noting the abrupt change of concentration along a straight line  displayed in figure  \ref{fig:Morpholohy}b for the loose  granular collapse. This line splits the moving  and the static regions, and it's commonly referred to as the failure line. The collapse of the upper right part of the column is triggered right after the gate is removed.  The sliding region is contracting more rapidly close to the bottom and the failure line because the shear rate is higher near the non-moving regions (the variable $\phi_{pl}$ embedding the dilatancy effects is proportional to the shear rate - see Eq. \ref{eq:volum_strain_temporal}). The failure line and the fluid velocity field are also visualized in figure \ref{fig:GradAlphaLoose}. As mentioned before, only the sliding wedge is moving significantly, hence contracting. This leads to an expansion along the failure line to ensure the conservation of mass. Figure \ref{fig:GradAlphaLoose}c illustrates the dilation rate as defined by \cite{iverson2014depth}, i.e. the divergence of the solid phase velocity ($\nabla \cdot \mathbf{u^s}$) so we can distinguish the contractancy and dilatancy regions. After the gate removal $\phi_{pl}$ increases along the failure line. According to Eq. \ref{eq:J_and_J2}, the contact pressure is reduced, therefore, a reduction of the shear strength is expected enhancing a rapid flow slide.  As the collapse carries on, the contractancy behavior of the sheared region entails an increment of the volume fraction, which according to Eq. \ref{eq:J_and_J2}, is accompanied by an increase of the contact pressure. Figures \ref{fig:Morpholohy}b and \ref{fig:GradAlphaLoose} also show a mild expansion within the non-moving region. One may wonder why the so-called non-moving region is deforming if we defined it as static region. It is worth mentioning that the expansion of the static region is mainly caused by the reduction of the column height and its subsequent decompression. As a matter of fact, the expansion rate of this region is significantly low  ($\nabla \cdot \mathbf{u^s}$ values in figure \ref{fig:GradAlphaLoose}c are $\nabla \cdot \mathbf{u^s} \approx  0.004 s^{-1}$ whereas   $\nabla \cdot \mathbf{u^s} \approx  -0.020 s^{-1}$ inside the contracting band). The gentle expansion of the static region is accompanied by an inward flux (see figure \ref{fig:GradAlphaLoose}c) that occupies the growing pore space.

Figure \ref{fig:Morpholohy} exhibits qualitatively  good agreement with the experimental data \citep{rondon2011granular} showing comparable time scales and remarkably similar run-out distances.  Although, there is an undeniable resemblance  between the numerical and the experimental final deposits, figure  \ref{fig:Morpholohy} suggests that  the numerical solution for the dense collapse leads to a milder final slope compared to the experimental data and the loose case is slightly slower during the first moments after the gate removal.

\subsubsection{Excess of pore pressure}

\begin{figure}
\centering
\includegraphics[width=0.99\textwidth]{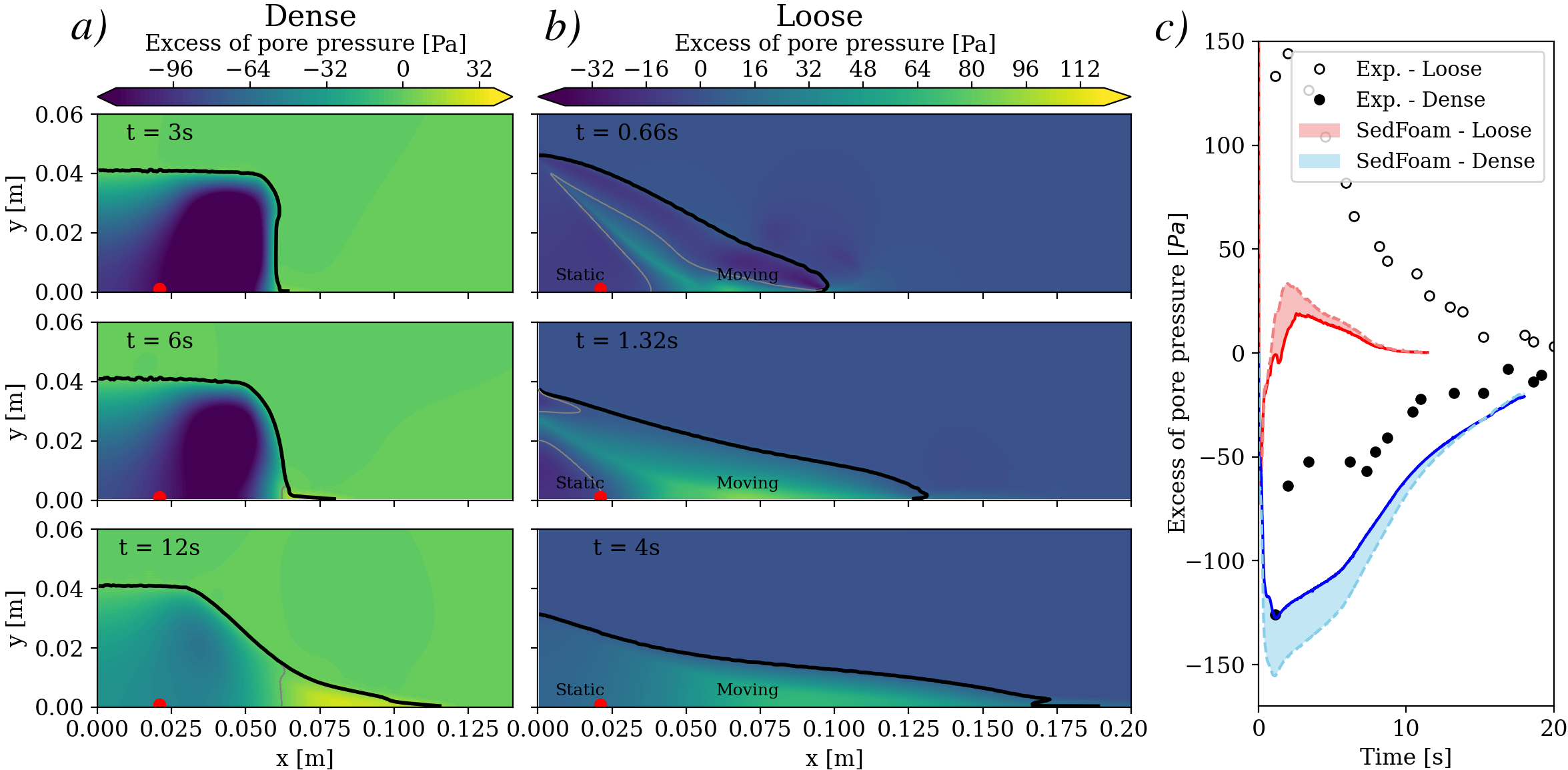}
\caption{\label{fig:PressureField}Evolution of the morphology and excess of pore pressure ($p^f$) during the collapse of an initially a) dense and b) loose column. A gray line is included to illustrate the zero pressure isoline. c) Evolution of basal pore pressure ($p^f$) measured at $2cm$ ($-$ dark continuous  line) and  $3cm$ ($--$ light dashed line). Shaded areas correspond to the region between the two probes results.}
\end{figure}

Figure \ref{fig:PressureField}a shows that negative pore pressure develops, stabilizing the dense granular material. The initial vertical front has a much steeper  slope than the angle of repose. Because of this unstable configuration, shearing deformation is triggered. Under such circumstances, the granular material dilates, pore bodies are enlarged and the fluid phase flows inwards the porous medium to accommodate the expansion. Consequently, negative pore pressure is generated increasing the effective strength, and overall, stabilizing the granular material. The negative pore pressure is, therefore, responsible for the characteristic creeping flow observed in figures \ref{fig:Morpholohy}a and \ref{fig:PressureField}a.

The loose scenario portrayed in figure \ref{fig:PressureField}b shows a dual positive/negative pore pressure map revealing the complex dynamics of the collapse.  The contracting behavior of the sheared region comes along with an expulsion of the pore fluid. Therefore, positive pore pressure develops within the moving area reducing the shear resistance and enhancing a rapid granular flow.   As mentioned before, dilation also occurs in the loose collapse along the failure line (see figure \ref{fig:GradAlphaLoose}c). Subsequently, the fluid is sucked into the failure line and partially into the non-moving region due to its decompression leading to negative pore pressures. This phenomenon was already reported by \cite{lee2021two}: the solid volume fraction along the failure line decreases inducing a reduction of the solid pressure (see Eq. \ref{eq:J_and_J2}), thus a lower shear-strength. As a result of the combination of positive pore pressure within the sliding zone and a significant decrease of contact forces along the failure line, the shearing region is partially fluidized and the lower effective stress is accompanied by a rapid slide failure. The pore pressure map depicted in figure \ref{fig:PressureField}b is consistent with the  pore pressure feedback mechanism and the sliding failure reported in the experiments. However, pore pressure values at the bottom of the loose column (see red point located at $2cm$ from the end of the tank in figure \ref{fig:PressureField}b) differ from  the experimental measurements.  Figure \ref{fig:PressureField}c shows the experimental data and the numerical solution have a similar trend for the loose and dense granular collapse: a pore pressure jump is registered after the gate removal that  gradually dissipates.  The negative pore pressure jump simulated for the dense granular collapse is properly predicted, however, the dissipation dynamics observed in figure  \ref{fig:PressureField}c are slightly different. Even though the failure mode and the pore pressure feedback mechanism are generally well reproduced, the positive pore pressure values are underpredicted for the loose case. It is worth noting that the model fails to replicate the exact position of the failure line, and therefore, the pore pressure probe (at $2cm$ from the end of the tank) falls into the non-moving area instead of the sliding region where pore pressures present two distinct behaviors: negative pore pressure within the static zone and positive pore pressure in the sliding region. For this reason, figure  \ref{fig:PressureField}c includes the pore pressure numerical measurements between $2$ and $3cm$.

\subsubsection{Sensitivity study}\label{sensitivity}

This section summarizes the role of different parameters on the dynamics of the granular collapse.   Figures on the sensitivity analysis  are provided as supplementary material. The results of the sensitivity study were used to find  the optimal set of  parameters to reproduce  Rondon's experiments numerically.

\paragraph{$K_1$ sensitivity}\label{KDilaSensi}

\noindent

The dilatancy prefactor $K_1$ introduced in Eq.\ref{eq:dilatancy} is responsible for controlling the plastic effects that arise from particle rearrangements during shear deformations. 
From the results shown in  the supplementary material for the dense case, we observe that large $K_1$ values result in slightly slower creeping flow while the final deposit shape is unaltered.  Dilatancy is relevant right after the gate removal, then, dilatancy effects fade out and the granular collapse starts flowing after  $t \approx 6s$ as it would flow at the equilibrium state ($\phi=\phi_{\infty}$), eventually, reaching the same deposit shape regardless of the  $K_1$ value. Numerical results show that the dense granular collapse follows very similar dynamics for both $K_1=4$ (value proposed in \cite{montella2021two}) and $K_1=40$ (reference value in this article). The $K_1=4$ scenario provides a slightly faster collapse but a better pore pressure dissipation curve matching the experimental points with striking accuracy.

Concerning the loose case,  supplementary material   illustrates increasing  $K_1$ has a strong effect on the dynamics of the spreading deposit. According to Eq.\ref{eq:volum_strain_temporal} and Eq.\ref{eq:dilatancy} , the increment of $K_1$ is responsible for $\phi_{pl}$ to increase more rapidly. Microscopically, it means contact forces are reduced more abruptly leading to a lower solid pressure and shear strength. Consequently, large $K_1$ values enhance a rapid flow with longer run-out distances.  Adopting $K_1=4$ for the loose case is not enough to trigger compacting effects and the corresponding pore pressure feedback, meanwhile, the largest value ($K_1=100$) reveals a high positive pore pressure jump. However, the morphology of the deposit remains  barely affected.  The lack of difference between $K_1=40$ and $K_1=100$ in terms of deposit shape for both loose and dense collapses is a consequence of the dilatancy coefficient ($\delta $) limits  imposed to keep  $\delta $ bounded to the physical values reported in \cite{pouliquen1996onset,iverson2014depth,alshibli2018influence}.  $\delta $ values are significantly important after the gate removal. At this point,  both  $K_1=40$ and $K_1=100$  scenarios reach the limit $| \delta  | =0.4$ in some regions of the granular column, thus, no relevant difference is observed between  $K_1=40$ and $K_1=100$.  As we approach the equilibrium state, $\delta $ values decay suppressing the dilatancy effect.

The results presented in this subsection indicate that, overall, good agreement is found between the experimental data and the numerical simulations.  However, it is worth mentioning that the choice of $K_1$ is strongly influenced by the reference critical volume fraction, which is discussed in the next subsection, therefore,   one may find other optimal $K_1$ values after increasing/decreasing the critical volume fraction $\phi_c$. This dense-loose asymmetrical trend in terms of $K_1$ sensitivity  suggests that the dilatancy model still has room for improvement. Indeed, the dilatancy model is governed by the evolution of the dilation angle through changes in the plastic volume fraction using the first term of the Taylor expansion of the dilatancy expression given by  Eq. \ref{eq:dilatancyDef}. Additionally, the elasto-plastic expression given by Eq. \ref{eq:J_and_J2} is a crude simplification to model the stress field neglecting the anisotropy and non-local effects of real soils. Thus, the current formulation is a simplified approach to model  dilatancy effects based on the amount of plastic volumetric strain. Further research could explore the use of non-linear dilatancy laws and perform DEM simulations to improve or introduce new phenomenological expressions to predict the effects of dilatancy with even greater accuracy.

\paragraph{$\phi_c$ sensitivity}\label{phiCSensi}

\noindent

The influence of the critical volume fraction ($\phi_c$) is closely linked to the dilatancy effects. According to Eq. \ref{eq:volum_strain_temporal} and Eq. \ref{eq:phi_viscous}, dilatancy effects are proportional to $\phi-\phi_c/(1+I_v^{1/2})$. Therefore, assuming lower $\phi_c$ values result in lower contractancy effects for initially loose cases and stronger dilatancy effects for initially dense packings (see figure in the supplementary material). Likewise, larger $\phi_c$ values are associated to weaker dilatancy in dense cases and enhanced contractancy for loose granular columns. Changes in $\phi_c$  are remarkably more important in the loose packings.  Contractancy effects partially fluidize the granular column, which leads to rapid collapse with a final deposit of a very gentle slope. Correspondingly, the pore pressure dynamics are significantly higher for the loose case. In particular, the $\phi_c=0.58$ case reproduces the magnitude of the positive pore pressure jump reported in the experiments. On the contrary, minor differences in terms of pore pressure are observed for the scenarios with initially dense columns.

\paragraph{$K_2$ sensitivity}\label{K2Sensi}

\noindent

In this section, we examine the impact of the parameter $K_2$ in Eq. \ref{eq:pss_relax}. $K_2$ affects how quickly the shear-induced pressure reaches its equilibrium state. Figures in the supplementary material demonstrate that there is barely no difference in terms of excess pore pressure and deposit spreading response. It is important to note that the numerical simulations in this study are conducted under a dense viscous  granular flow regime. Therefore, it is not surprising that the contact pressure has a much greater impact compared to the shear-induced pressure. In the dense scenario $K_2=1$ poses numerical stability issues that require a smaller time step. Thus,  $K_2=0.01$ is preferred for the dense granular collapse to circumvent numerical instabilities.

\paragraph{Elastic modulus}\label{ESensi}

\noindent

This section explores the influence of the  elastic modulus ($E$) of Eq. \ref{eq:J_and_J2} on the granular collapse dynamics. Before getting into the discussion, it is pertinent to note that $E$ values used in the present numerical model are considerably far from real materials such as glass beads ($E \approx 70GPa$) or sand ($E \in  [5-80]  MPa$), however, these values would induce numerical issues with the non-linear approach of Eq. \ref{eq:J_and_J2}. Instead, $E$ values used in the present sensitivity analysis remain in a lower range ($E \in  [0.1-100]  Pa$) where numerical instabilities are not detected. Limited differences are observed for the dense granular collapse in terms of pore pressure and deposit shape. However, results are more sensitive in the loose scenario. As detailed in the supplementary material, differences arise as a consequence of changes on the shape of the vertical concentration profile. The nature of the contact pressure expression (Eq. \ref{eq:J_and_J2}) plays a key role on the distribution of the initial volume fraction along the vertical by increasing the concentration vertical gradient with soft elastic modulus. A low elastic modulus (i.e. $E=0.1Pa$) leads to a vertical concentration curve that ranges from $\phi_{top}=0.525$ to $\phi_{bottom}=0.565$ ($\Delta  \phi= \phi_{bottom} - \phi_{top} \approx 0.04$) a whereas stiffer modulus  (i.e. $E=100Pa$) has a narrower range ($\Delta  \phi= \phi_{bottom} - \phi_{top} \approx 0.01$). Choosing a low elastic modulus leads to a dual behavior within the granular column; the material close to the bottom  shows the classic features of a dense soil while the material located close to the surface presents a  very loose-like behavior. It is, therefore, recommended to use a high elastic modulus, provided that the numerical model remains stable, in order to have a realistic soil behavior.

\paragraph{Permeability coefficients}\label{other}
\noindent

Figures in the supplementary material show lower permeabilities lead to a slow mobilization. In turns, the pressure dissipation takes longer as expected, specially for the initially dense column. Figures in the supplementary material evidence different Engelund's coefficients have a minor impact on the results for the loose granular column regarding the morphology and the pore pressure curve.

\paragraph{Frictional coefficients}\label{Frictional}
\noindent

The friction coefficient (see Eq.\ref{eq:muIVEq}) has a certain effect on the shape of the  deposit. Large friction coefficients delay the collapse and the deposit ends up with a steeper slope. The lower mobilization entails a weaker pore pressure feedback: the soil is more difficult to shear, thus, pore volume changes take longer.  Conversely, low fiction angles promote a rapid failure with abrupt pore volume changes, therefore, higher pore pressure jumps are observed.

\paragraph{Discussion on the sensitivity study}\label{disc}
\noindent

In this work we have optimized the set of parameters  based on the shape of the deposit. The sensitivity analysis of the present  section reveals that a different  set ($\phi_c \uparrow$, $K_1 \uparrow$ and $\mu_s \downarrow$) would definitely provide a better pore pressure prediction for the loose case in detriment of the prediction of the  deposit morphology. Nevertheless, none of the combinations (except for $\phi_c=0.58$) presented in the sensitivity analysis  reaches the positive pore pressure developed within the loose granular column. Several factors could explain the underprediction of the positive pore pressure curve: 1) results are very sensitive to the initial volume fraction. Indeed,   \cite{lee2021two} reported  different $p^f - time$ curves and deposit shapes for a narrow range of initial concentrations: $\phi_o = 0.553$ and  $\phi_o = 0.550$. 2)  Measurement imprecisions and external factors inherent to the experiment may not be completely modeled in the numerical simulations. For instance, free surface perturbations could be induced at the moment of gate removal and/or rapid collapse of the loose column. Such perturbations may increase the pore pressure and partially fluidize the granular column.  \cite{rondon2011granular} claimed that  wall effects are negligible, however, the fluid flowing through the porous medium may have three-dimensional effects that can't be captured by the 2D numerical approach. 3) Limitations of the numerical model to fully reproduce the pore pressure feedback mechanism for a wide range of concentrations using a single formulation and set of numerical parameters.

\subsection{Breaching process}

\subsubsection{Experimental set-up}

Unlike loose shear failures, where the collapse is almost instantaneous, initially  dense packings  are distinguished by their slow collapse and the presence of negative  pore pressures that stabilize the soil. At larger scales, it is also possible to observe the breaching process. As described in section \ref{RondonSect},  dilatancy induces negative pore pressure that reduce the shearing preventing a shear failure. However, near the front face, particles are released due to the expansion of the granular material causing the breach front to slowly regress.

\begin{figure}
\centering
\includegraphics[width=0.9\textwidth]{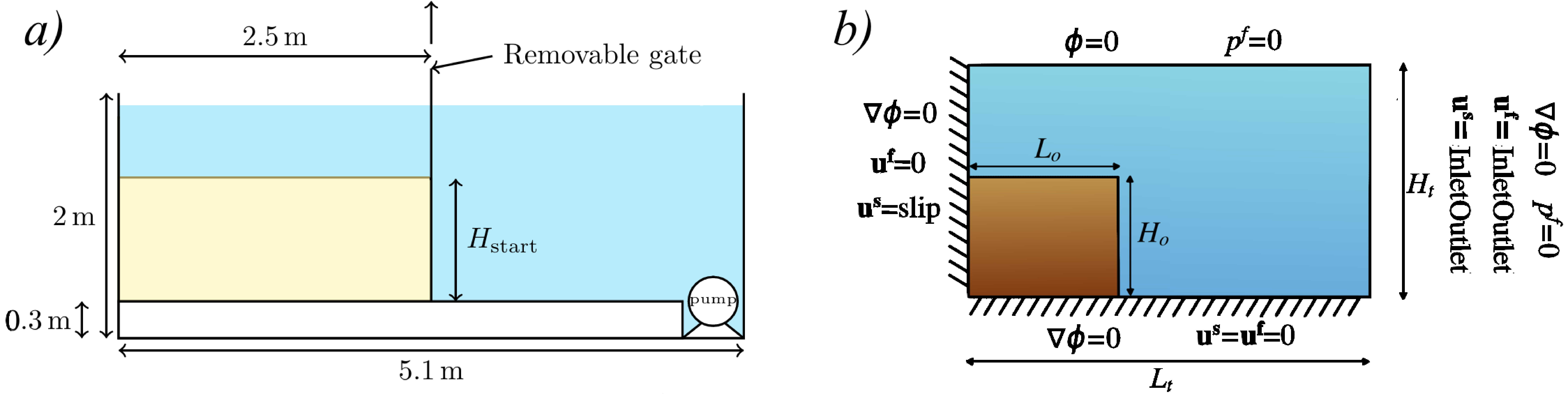}
\caption{\label{fig:SchemeWeij} a) Experimental set-up to study the breaching process. Image taken from  \cite{weij2020modelling}. b) Numerical setup. In SedFoam, the "InletOutlet" condition is written as "pressureInletOutletVelocity" so the velocity is set   to have a zero gradient condition when the flow leaves the domain, whereas, the velocity assigned  when  the flow goes into the domain is based on the flux in the patch-normal direction. }
\end{figure}

In this section, the laboratory experimental data provided by \cite{weij2020modelling,alhaddad2023stabilizing} are used to evaluate the accuracy of the present numerical model to predict the breaching process. The experimental setup of \cite{weij2020modelling,alhaddad2023stabilizing} consists of a tank with a height of 2 meters, a length
of 5.1 meters and a width of 0.5 meters (see figure \ref{fig:SchemeWeij}). An impermeable removable gate, similar to the one used in Rondon's experiments, is placed at a distance of 2.5 meters to divide the tank in two sections. A pump is located at the bottom right corner as shown in figure \ref{fig:SchemeWeij}. The location of the pump prevents the reflection of the turbidity current.  This pump takes out the mixture to a basin which is 4.5 m long, 1.25 m wide, and 1.25 m high. In addition to this pump, a second pump is placed  behind a 1 m high divider to  maintain a constant water level by pumping clean water back into the reservoir. The left part of the tank is filled with sand prepared by the following procedure: the sand is placed horizontally layer by layer. In order to achieve a dense granular packing, layers are compacted using a vibrating needle. This process goes on until the height of the sand column is 1.47 meters.

Two different types of sand are considered in the laboratory experiments: 1) GEBA sand with a median diameter of $D_{50}=120 \mu m$, initial volume fraction of  $\phi_o=0.585$, and an internal friction angle of $\mu_s=35.8^o$ and 2) D9 sand with a median diameter of $D_{50}=330 \mu m$, initial volume fraction of  $\phi_o=0.570$, and an internal friction angle of $\mu_s=40.1^o$. For the numerical simulations, we observed that the size of the mesh elements should be at least $\Delta x =2.5 mm$ (a uniform grid is considered). Larger grid sizes accelerate the breaching and produce more rounded shapes at the corner of the top right column are found. Although the mesh convergence is not completely reached, the computational cost becomes too expensive for finer meshes with little effect on the results in terms of morphology and wall velocity
(a mesh convergence study has been carried out and it is available in the supplementary material). Thus, the choice of $\Delta x =2.5 mm$ seems reasonable to study the problem  without compromising  the accuracy of the solution.  As shown in section \ref{phiCSensi} and suggested by \cite{weij2020modelling},  the critical volume fraction $\phi_c$ controls the amount of dilation, hence, the velocity of the receding wall. Accordingly to \cite{weij2020modelling}, the critical volume fraction is chosen be the same  as the sand concentration during breaching, just before it is released from the breach face. In this case,  $\phi_c=0.545$ and $\phi_c=0.56$ are chosen for the GEBA and D9 sand, respectively.  Finally,  the Engelund's coefficients (linked to the permeability of the soil) are chosen as $\alpha_E=1200$ and $\beta_E=3.6$ for the  GEBA sand and $\alpha_E=900$ and $\beta_E=1.8$ for the D9 sand. These values have been calibrated to match the experimental wall velocity (the horizontal velocity at which the steep slope moves  due to the breaching process). Rheological and numerical parameters are summarized in table \ref{tab:table_NumWeij}. Numerical results on the Rondon's experiment (see section \ref{RondonSect}) evidenced small differences between $K_1=4$ and $K_1=40$. In this section we take $K_1=4$ because most of the works in literature adopt  dilatancy factors with the same order of magnitude.

\begin{table}
  \begin{center}
\def~{\hphantom{o}}
  \begin{tabular}{cccccccccccccc}
      \textbf{Parameter}& $H_t$ & $H_o$ & $L_o$ & $L_t$ &   $\mu_s$ &  $\Delta \mu$   & $I_o$ & $K_1$ & $K_2$ & $\phi_c$ & $\alpha_E$ & $\beta_E$   \\[2pt]
  \textbf{GEBA sand}& 2.0 & 1.47 & 2.5 & 5 & 0.72 & 0.34 & 0.004  & 4 &  0.01  & 0.545 &  1200 &  3.6\\
    \textbf{D9 sand}& 2.0 & 1.47 & 2.5 & 5 & 0.84 & 0.34 & 0.004  & 4 &  0.01 & 0.560 &  900 &  1.8\\

  \end{tabular}
  \caption{Geometric, rheological and numerical parameters used to reproduce \cite{weij2020modelling}.}
  \label{tab:table_NumWeij}
  \end{center}
\end{table}

At this point, its is worth noting that  Eq. \ref{eq:mom_solid} and \ref{eq:mom_fluid} neglect the contribution of the turbulent suspension term. One may argue that at the breach face a turbidity current may be triggered, in which case, turbulent suspension/dispersion could significantly contribute to the erosion and this would need to be modeled through a turbulent suspension term in the momentum equation. However, this process takes place at a length scale which is on the order of the grain-scale,   smaller than the grid size  of our numerical simulations, therefore, it cannot be resolved by the two-phase flow model in the experimental breaching configuration. In addition to the smaller grid sizes at the interface required to capture the transition from regions dominated by the granular rheology to the areas where the turbulent suspension term is dominant,  other missing multiphase forces in our model, such as the lift-force near the interface, may change the turbulent kinetic energy  and affect the transition towards the turbulence suspension. Numerical tests with the turbulence term switched on have been performed showing an overestimated wall velocity as well as a complete erosion of the bottom deposit. The rapid erosion is probably due to the inability of the model to capture the correct turbulent suspension term. This term is proportional to the gradient of the solid volume fraction which becomes significantly high at the interface, where the volume fraction changes from $ \sim 0$ to $ \sim 0.6$ over a very short distance. As the capability of the two-phase model to simulate turbulent erosion by turbidity currents has not been checked even in simpler idealized situations, this would deserve further investigation that is beyond the scope of the present paper.

\subsubsection{Dynamics of breaching}

The removal of the gate triggers an initial creeping phase. The low mobility of the column is associated to the negative pore pressure measured with the sensors. Depending on the type of sand, different dynamics are observed.

\begin{figure}
\centering
\includegraphics[width=1.0\textwidth]{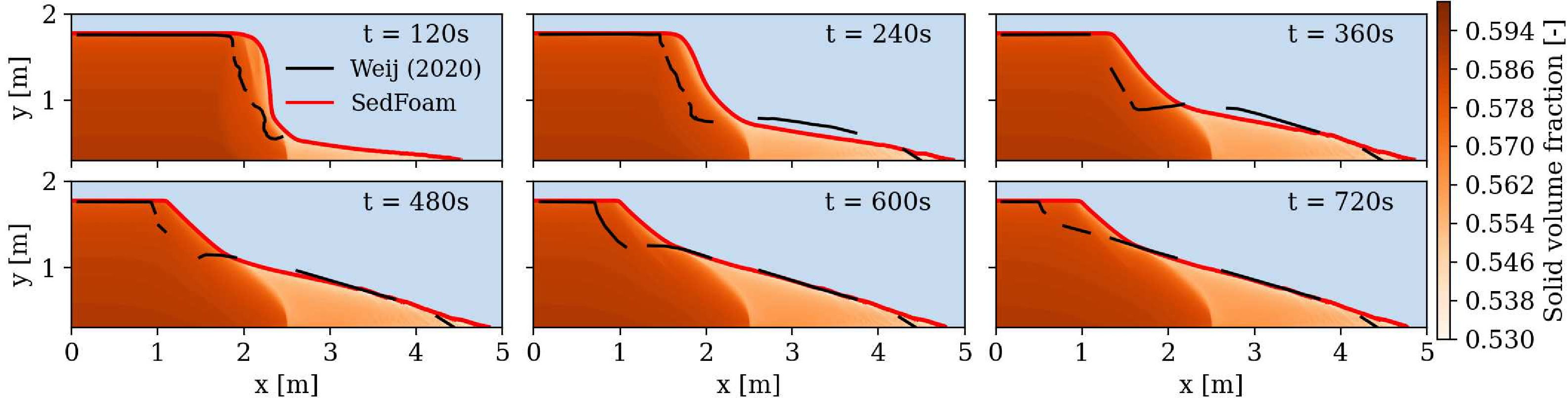}
\caption{\label{fig:Weij_Exp8} Comparison of the morphology  between the experiments and the numerical simulations for the GEBA sand. }
\end{figure}

In figure \ref{fig:Weij_Exp8} we observe the evolution of the column morphology for the GEBA sand. The velocity of the receding front is well captured by the numerical model, however, slight discrepancies are found near the right top corner, where the numerical model predicts a  rounded shape rather than a sharp corner observed in the experiments. The final shape consists of a deposit with a slope around $20^o$  modeled by SedFoam with remarkable agreement.

\begin{figure}
\centering
\includegraphics[width=0.99\textwidth]{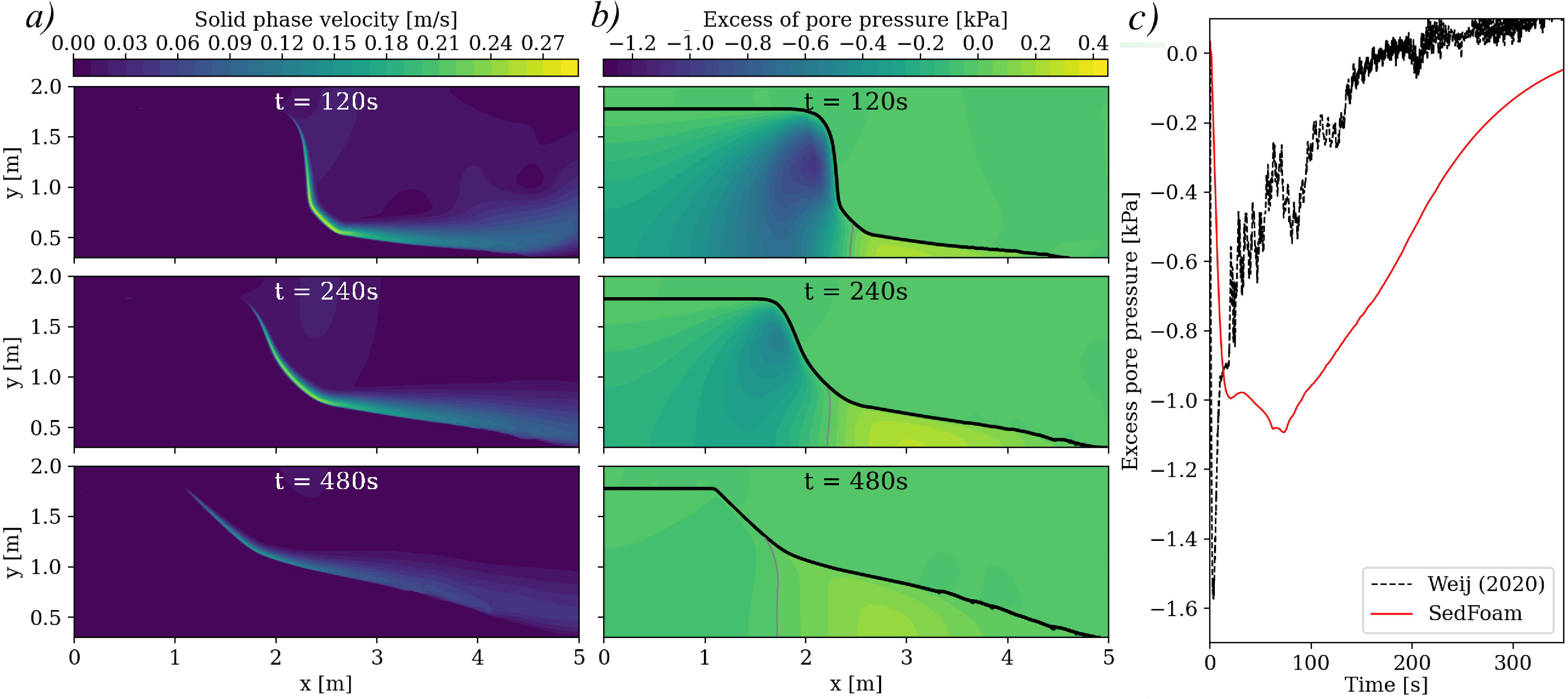}
\caption{\label{fig:Weij_Exp8_porepressureVel} a) Solid phase velocity field and b) pore pressure field extracted in the numerical simulations for the GEBA sand.  A gray line is included to illustrate the zero pressure isoline. c) Comparison of the excess of pore pressure ($p^f$) evolution within the granular column  between the experiments and the numerical simulations for the GEBA sand. }
\end{figure}

A key feature of the breaching process is the turbidity current formed near the front face as the particles are slowly released and pulled down by gravity. Such currents are illustrated in figure \ref{fig:Weij_Exp8_porepressureVel}a and become less intense with time. Some particles released from the breaching front settle and contribute to form the deposit at the bottom. Although the main physics of turbidity currents are well captured, the lack of the turbulent suspension term may introduce inaccuracies in our numerical results. The presence of the deposit reduces the height of the breaching face and, by the end of the experiments and the numerical simulation, the deposit adopts  a triangular shape with a slope milder than the friction angle in which the material is at rest with no turbidity current.

The negative  pore pressure plays the role of stabilizing the column and contributes to delay the granular flow. Figure \ref{fig:Weij_Exp8_porepressureVel}b shows a pore pressure map very similar to the  Rondon's granular collapse studied in section \ref{RondonSect}.  The negative pore pressure is significantly higher in the right top region of the granular column, where the material is more likely to fail along a shear plane. On the one hand, the negative pore pressure is responsible for a higher shear strength that keeps the granular column as a whole without shear failures. On the other hand, the dense material at the top right area is expanding which enhances the turbidity currents previously discussed. As the pore pressure dissipates, the shear resistance is reduced and the shear stress may reach the yield point forming a shear failure plane. This situation was reported in a few   experiments \citep{weij2020modelling} where occasional shear slide failures were observed during the breaching process. When slides occur, small drops in the excess of pore pressure are measured by the pressure sensors near the shear plane. In particular, the pressure drops located at $t \approx 80s $ and $t \approx 125s $  could have caused two  successive minor slides in the experiments as shown in figure \ref{fig:Weij_Exp8_porepressureVel}c.  These slides erode the front face much faster than the breaching process. Nonetheless, no sliding failures were observed in the numerical situations for the GEBA sand where the receding front face was mainly conducted by the breaching process. In the numerical simulations, figure \ref{fig:Weij_Exp8_porepressureVel} shows a pore pressure drop right after the gate removal followed by a continuous dissipation with a time scale and dissipation rate comparable with the experimental ones.

\begin{figure}
\centering
\includegraphics[width=0.8\textwidth]{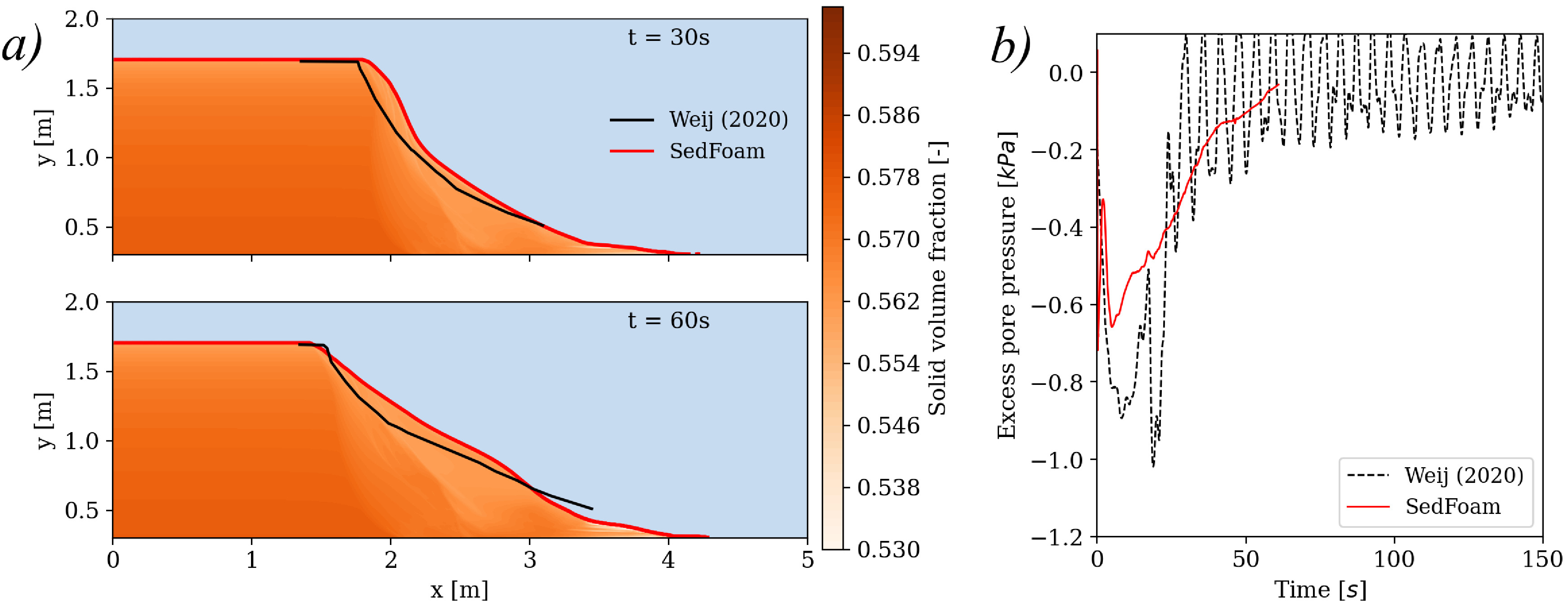}
\caption{\label{fig:Weij_Exp16} a)  Comparison of the morphology and b) the pore pressure ($p^f$) evolution within the granular column  between the experiments and the numerical simulations for the D9 sand.}
\end{figure}

The experiment and numerical simulations using the D9 sand show a similar trend, yet, slightly different dynamics are observed in the experiments.  The larger permeability of the D9 sand (grains roughly three times larger than the GEBA sand and lower initial volume fraction) leads to a faster breaching process. In figure \ref{fig:Weij_Exp16}a  the evolution of the deposit morphology is well predicted by the numerical model in terms of shape and time scale. Despite the good agreement observed in figure \ref{fig:Weij_Exp16}a,  in the experiments, the initial breaching process is followed by a slow sliding failure. This feature is not captured by the numerical model.  The \textit{dual mode failure}, including the breaching process  and punctual shear failure, is  evidenced in figure \ref{fig:Weij_Exp16}b. Initially, the negative pore pressure builds up within the granular material stabilizing the whole granular column. Then, the progressive dilation of the medium is accompanied by a reduction of the pore pressure. At $t \approx 18s$, the negative pore pressure is no longer capable to counteract the shear forces, therefore, a wedge close to the breaching face starts sliding down. Simultaneously, the sliding region increases the shear rate significantly, especially near the failure line, which is accompanied by a negative pore pressure build up (second minimum point in \ref{fig:Weij_Exp16}b). In contrast to the experimental measurements, the numerical approach is unable to reproduce this particular   \textit{dual mode failure}.

\section{Conclusion}\label{Conclusion}

The present article studied the sliding and breaching failures for different granular columns. In a first series of numerical tests, the model showed to be able to reproduce
the experimental data of  \cite{rondon2011granular}  with a reasonable agreement. We showed that dilatancy effects are crucial to reproduce the rapid collapse of initially loose columns and the low mobility of initially dense columns. Additionally, the  plastic effects embedded in the present dilatancy model are a key feature to predict the consequent pore pressure feedback mechanism signified by negative pore pressure for the dense case and positive pore pressure in the loose columns.  The results presented in the sensitivity analysis suggest that dilatancy effects intensify with large $K_1$ values or with the choice of $\phi_c$ values far from the actual volume fraction. Concerning the basal  pore pressure, little sensitivity to the dilatancy effects ($K_1$ and $\phi_c$) is found for the initially dense column. However, the influence of dilatancy effects is much stronger for the loose case. Indeed, large $K_1$ or $\phi_c$ values lead to a partially fluidized bed. In such cases, some regions of the column exhibit high positive pore pressures that counterbalance the gravitational forces and the mixture flows easily until the pore pressure has dissipated so the shear strength builds up again. Although the model provided fairly good agreement with the experiments capturing the dilatancy effects at the grain scale through the dilatancy model proposed by \cite{montella2021two} and the expansion/contraction of the granular medium governed by $\phi(I)$ (see Eq. \ref{eq:phi_viscous}), improvements on the dilatancy model could reduce and explain the current discrepancies, in particular, the underprediced positive pore pressures observed during the loose collapse. On the one hand, other dilatancy laws beyond the linear expression given by Eq. \ref{eq:dilatancy}
should be explored. On the other hand, non-local dynamics should be introduced in the model to take into account the system size, which in turns, has an important  effect on the level of stresses as reported by  \cite{athani2022transient}.

For the second experimental configuration, the breaching process of  \cite{weij2020modelling} was numerically reproduced with very good agreement. The present numerical model was able to reproduce the breaching process with great success in terms of morphology and  a reasonably good prediction of the pore pressure measured within the granular medium. \cite{weij2020modelling} experiments evidenced that in some cases, sliding failures occur in addition to  the breaching process. Although, \cite{weij2020modelling} did not report  sliding failures in most of the experiments, they suggested  the \textit{dual mode failure} is rather common in nature. Our numerical simulation mainly reproduces breaching but different set of parameters may predict features observed in slide failures. It seems, therefore, reasonable to include the transition of slope failures from breaching to slides in future  research in order to make an accurate prediction on the slope failure modes. As a main conclusion, it has been shown that a two-phase flow numerical model including dilatancy effects is able to reproduce the various failure modes of underwater particles collapse, breaching and slide failures, and their sensitivities to the initial dense volume fractions as well as their dynamics and morphology of the final deposits. The results of the present model could be extended in the future to investigate the influence of several geometrical and physical parameters on the dynamics of granular collapse without the need of experiments. More particularly,  the dredging industry could benefit from the present two-phase model to predict the flow slides during the excavation process, thereby further action may be considered to mitigate the erosion process. Finally, the present model could  shed some light on the transition between breaching and liquefaction flow slides.

\section{Acknowledgements}\label{Acknowledgement}
The authors  also would like to acknowledge the Strategic Environmental Research and Development Program, United States (MR20-1478). We would also like to acknowledge the support from the US National Science Foundation (CMMI-2050854). Most of the computations presented in this paper were performed using the GENCI infrastructure under Allocations A0060107567 and A0080107567 and the GRICAD infrastructure. Finally, we are also grateful to the developers involved in OpenFOAM.

\section{Data availability}\label{Data}
Supplementary material is found at the appendix. The numerical simulations corresponding to the  \cite{rondon2011granular} configuration  are available on  \footnote{https://github.com/SedFoam/sedfoam/tree/master/tutorials/laminar/2DCollapse}{\href{https://github.com/SedFoam/sedfoam/tree/master/tutorials/laminar/2DCollapse}{GitHub}}.  Results, post-process scripts and numerical setups for the breaching cases \citep{weij2020modelling} are available for download on \footnote{https://zenodo.org/record/8116463}{\href{https://zenodo.org/record/8116463}{
Zenodo}}.

\section{Funding statement}\label{Funding}
This work received funds from the Strategic Environmental Research and Development Program, United States, through the
grant number MR20-1478.

\section{Declaration of interests}\label{Declaration}
The authors declare no conflict of interest.

\section{Ethical standards}\label{Ethical}
The research meets all ethical guidelines.

\newpage

\appendix

\section{Supplementary material: Sensitivity analysis}\label{sensitibityAnnex}

The present sensitivity analysis provides a better understanding of the mechanisms that may affect the dynamics of the granular column collapse when some of the soil properties or numerical parameters are deliberately modified.

\subsection{$K_1$ sensitivity}\label{KdilaSensiSensiAnnex}

\begin{figure}[H]
\centering
\includegraphics[width=0.95\textwidth]{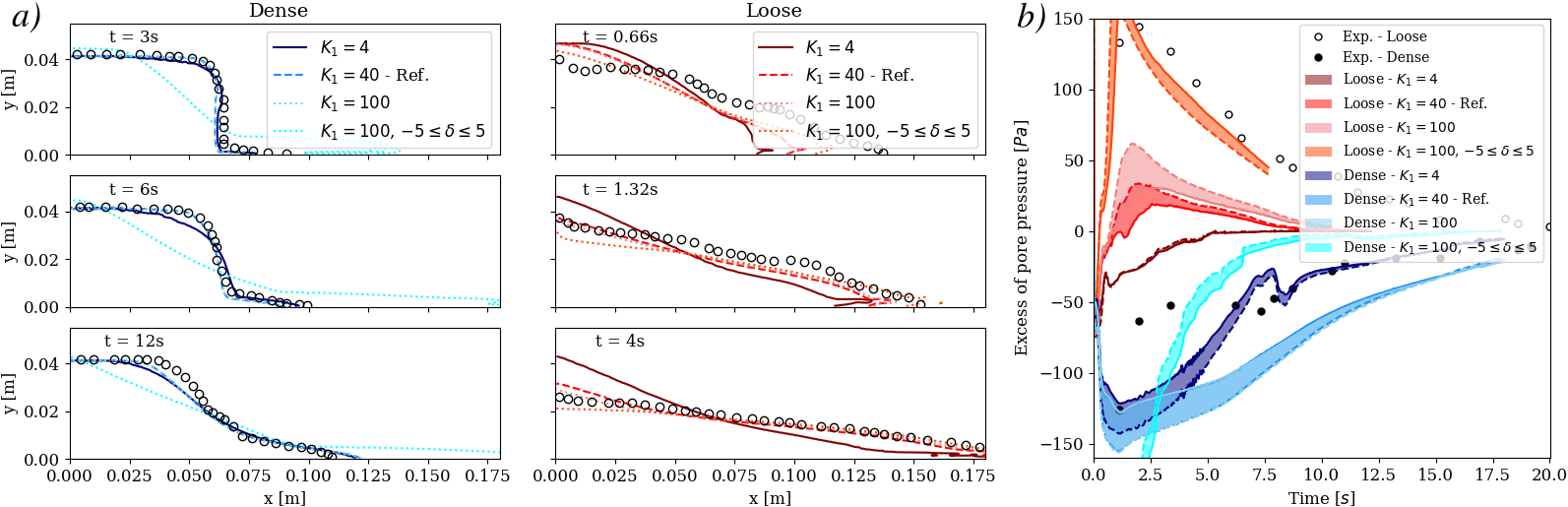}
\caption{\label{fig:Sensi_Kdila}Evolution of the morphology during the collapse of an initially dense and loose columns. b) Evolution of basal pore pressure measured at $2cm$ ($-$ continuous line) and  $3cm$ ($--$ dashed line). Shaded areas correspond to the region between the two probes results. The influence of $K_1$ is investigated.}
\end{figure}

Figure \ref{fig:Sensi_Kdila} displays the influence of  $K_1$ values on the collapse dynamics. We observe relatively small differences between   $K_1=40$ and $K_1=100$ due to the dilatancy coefficient ($\delta $) restricted to the  range  $-0.4 \leq \delta \leq 0.4$. After the gate removal,  both  $K_1=40$ and $K_1=100$  scenarios reach the limit $| \delta  | =0.4$ in some regions of the granular column, thus, no relevant difference is observed between  $K_1=40$ and $K_1=100$.  As we approach the equilibrium state, $\delta $ values decay suppressing the dilatancy effect, though in the $K_1=100$ case, dilatancy effects are mildly higher during the $\phi \rightarrow \phi_{\infty}$ transition.  

The loose case displayed in figure \ref{fig:Sensi_Kdila} shows $K_1$ has a stronger effect on the dynamics of the spreading deposit. Large $K_1$ values enhance a rapid flow with longer run-out distances because positive pore pressure develops within the granular medium (see the positive pressure jumps in figure \ref{fig:Sensi_Kdila}) reducing the shear strength of the system.

Figure \ref{fig:Sensi_Kdila}   also shows a case where the limit of the dilatancy coefficient  is set to  $| \delta  | =5$. This scenario predicts the correct positive overshoot of   pore pressure for the initially loose column but dilatancy effects fluidize a significant region of the granular column, which leads to a deposit shape that resembles a horizontal sediment layer. Regarding the dense case, the negative pore pressure is overpredicted, the expansion of the granular column occurs extremely rapid, so dilatancy effects go from an immediate large impact to an irrelevant contribution in the early stages. This results in a granular collapse that flows normally without dilatancy effects.
At this point, it is worth noting that in most of literature \citep{pailha2009two,mutabaruka2014initiation}, the dilatancy prefactor $K_1$ is  found to be in the range of  $K_1 \in [1 - 50]$.

One may think, following the logic of the dilatancy model presented in this work, that the pore pressure jump is higher because we are enhancing a rapid granular flow accompanied with contractancy, thus, a reduction of the pore space that results in positive pore pressure. However, the pressure jump is not only influenced by this transition, instead, the dynamics of the sliding failure and spreading are even more important. The fact that the shear strength is reduced more rapidly with increasing $K_1$ displaces the failure line quickly and the non-moving zone (where zero or negative pore pressures are observed - see figure \ref{fig:PressureField}b) is getting smaller much faster. The consequences of this mechanism are a more elongated bed, as more granular material is mobilized, and positive pore pressures measured at the probe location which is now placed under the sliding region.

\subsection{$\phi_c$ sensitivity}\label{phiCSensiSensiAnnex}

\begin{figure}
\centering
\includegraphics[width=0.95\textwidth]{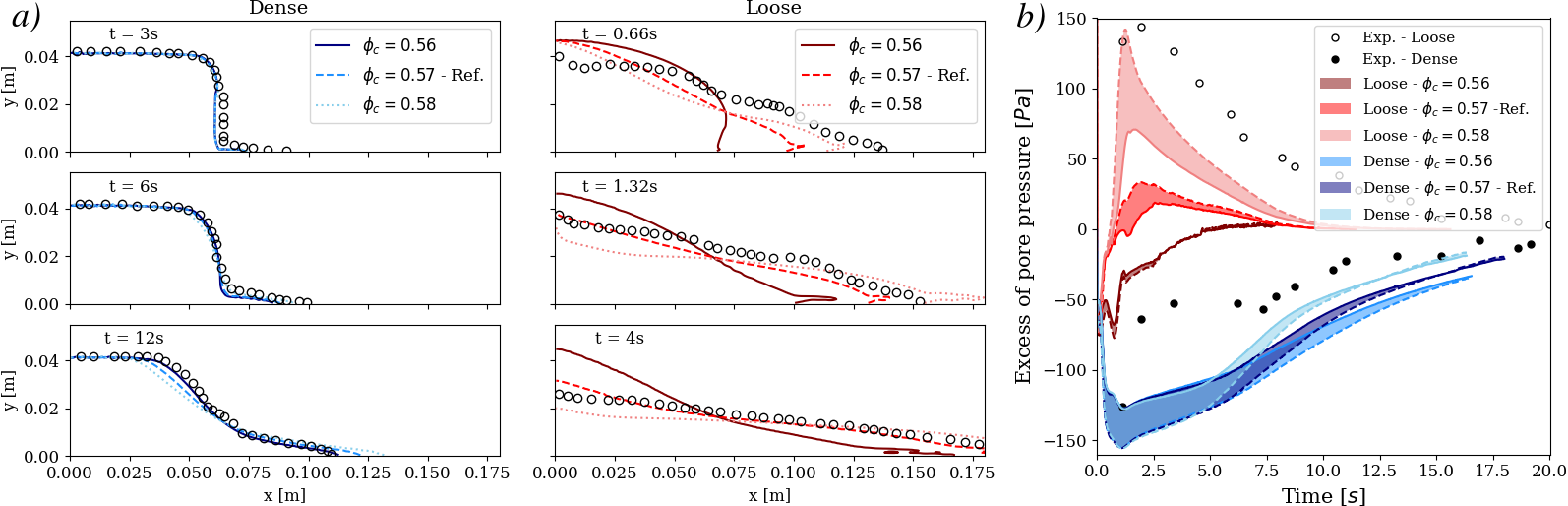}
\caption{\label{fig:Sensi_phic}Evolution of the morphology during the collapse of an initially dense and loose columns. b) Evolution of basal pore pressure measured at $2cm$ ($-$ continuous line) and  $3cm$ ($--$ dashed line). Shaded areas correspond to the region between the two probes results. The influence of $\phi_c$ is investigated.}
\end{figure}

The influence of the critical volume fraction ($\phi_c$) is displayed in figure \ref{fig:Sensi_phic}. By changing the value of the critical volume fraction one can adjust the 
dilatancy effects. Eq. \ref{eq:dilatancy} and Eq. \ref{eq:phi_viscous} show dilatancy effects are proportional to $\phi-\phi_c/(1+I_v^{1/2})$. Therefore, as shown in figure \ref{fig:Sensi_phic},  lower $\phi_c$ values result in lower contractancy effects for initially loose cases (the gap $\phi-\phi_c/(1+I_v^{1/2})$ is reduced) and stronger dilatancy effects for initially dense packings (the gap $\phi-\phi_c/(1+I_v^{1/2})$ is increased). Likewise, larger $\phi_c$ values are associated to weaker dilatancy in dense cases and enhanced contractancy for loose granular columns. Changes in $\phi_c$  are remarkably more important in the loose packings. Indeed, contractancy effects can fluidize a significant region of the granular column, which leads to a deposit shape closer to a horizontal sediment layer. Correspondingly, the pore pressure dynamics are significantly higher for the loose case. In particular, the $\phi_c=0.58$ case reproduces the magnitude of the positive pore pressure jump reported in the experiments. On the contrary, minor differences in terms of pore pressure are observed for initially dense columns.

\subsection{$K_2$ sensitivity}\label{K2SensiSensiAnnex}

\begin{figure}
\centering
\includegraphics[width=0.95\textwidth]{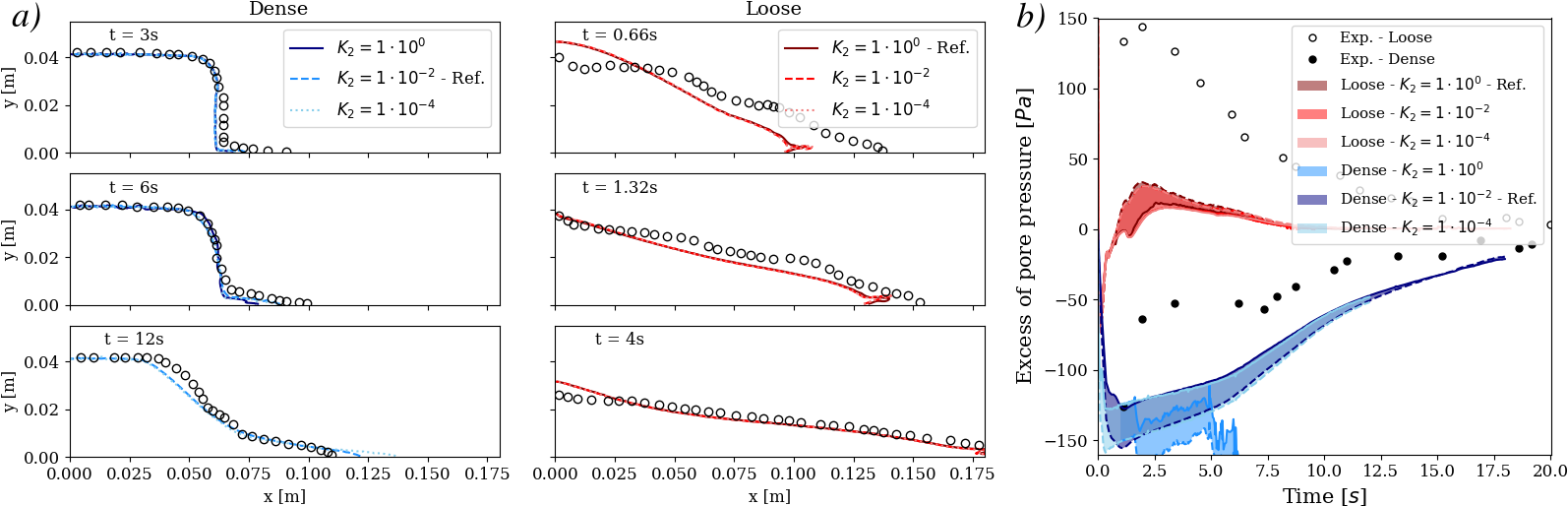}
\caption{\label{fig:Sensi_k2}Evolution of the morphology during the collapse of an initially dense and loose columns. b) Evolution of basal pore pressure measured at $2cm$ ($-$ continuous line) and  $3cm$ ($--$ dashed line). Shaded areas correspond to the region between the two probes results. The influence of $K_2$ is investigated.}
\end{figure}

Figure \ref{fig:Sensi_k2} illustrates the negligible effect of $K_2$ on the dynamics of the granular column collapse.  The shear-induced pressure remains residual and the collapse is controlled by  changes in the contact pressure.  Besides, the dense collapse simulation conducted using $K_2=1$ evidences undesirable numerical fluctuations that compromise the accuracy of the results, thus, additional relaxation is required to complete the numerical simulations with satisfactory results.  

\subsection{Influence of elastic modulus}\label{ESensiSensiAnnex}

\begin{figure}
\centering
\includegraphics[width=0.95\textwidth]{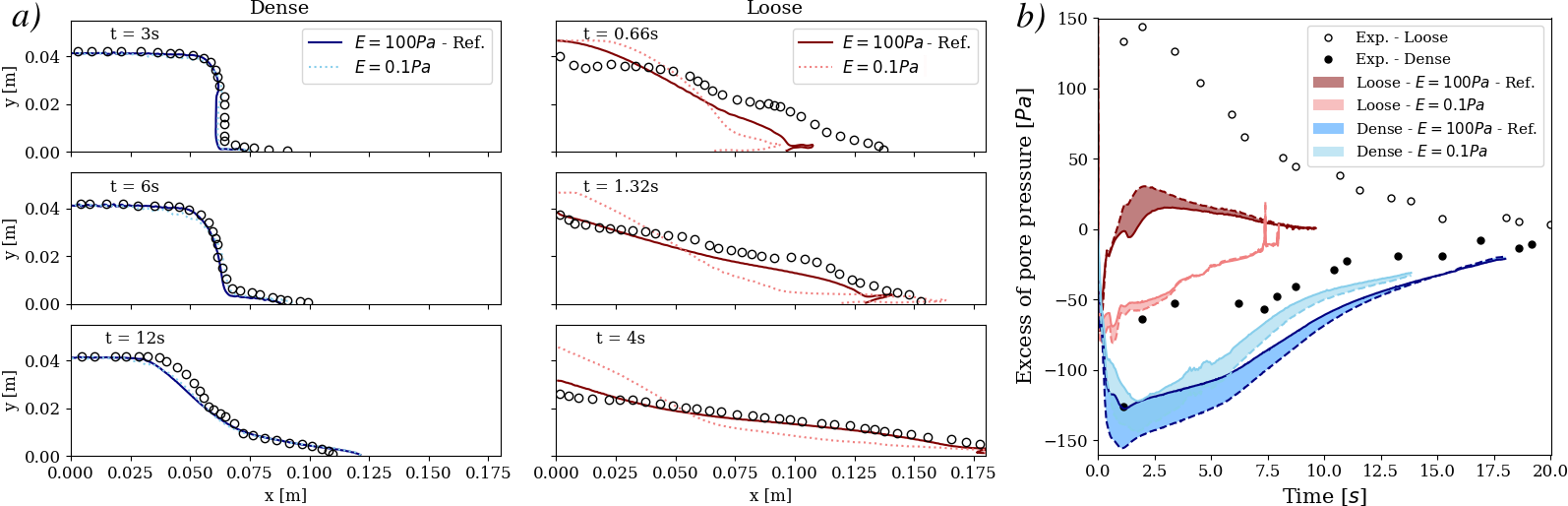}
\caption{\label{fig:Sensi_E}Evolution of the morphology during the collapse of an initially dense and loose columns. b) Evolution of basal pore pressure measured at $2cm$ ($-$ continuous line) and  $3cm$ ($--$ dashed line). Shaded areas correspond to the region between the two probes results. The influence of $E$ is investigated.}
\end{figure}

\begin{figure}
\centering
\includegraphics[width=0.5\textwidth]{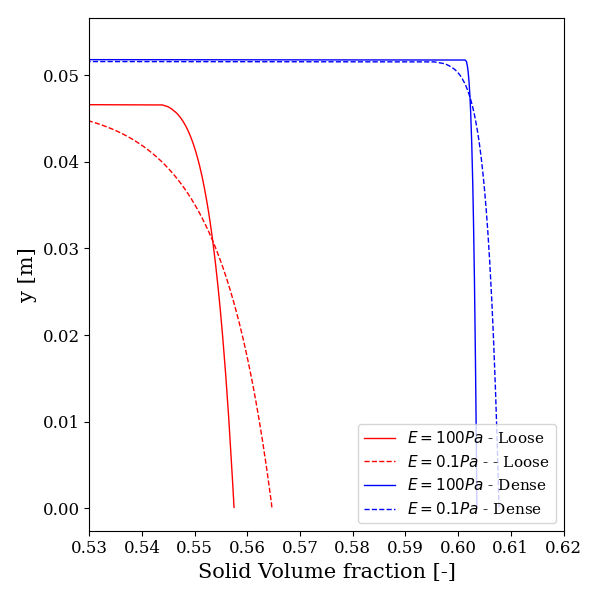}
\caption{\label{fig:Sensi_E2}Vertical concentrations profiles at the middle of the granular column ($x=L_o/2$) using different elastic modulus.  }
\end{figure}

Figure \ref{fig:Sensi_E} shows profiles and pore pressure curves for two different elastic modulus ($E$) exhibiting limited differences. However, these  differences are not solely due to the stiffness of the soil, but rather the different vertical concentration profiles that result from using various elastic modulus. Indeed, figure \ref{fig:Sensi_E2} provides evidence that using different elastic modulus values leads to different initial concentration profiles after the sedimentation.

The solid volume fraction is determined after the sedimentation process, where granular material settles under its own weight. Since we use an expression to model the granular pressure based on the volume fraction (see Eq. \ref{eq:J_and_J2}), once the sedimentation is complete, the weight of the granular layer is balanced by the contact pressure. This implies that we obtain a lithostatic granular pressure, which can only be achieved with a non-constant volume fraction profile. High values of $E$ result in steep concentration profiles, which are more realistic. However, numerical instabilities arise when $E$ exceeds $100 Pa$.
It is worth noting that for $E=100 Pa$, very small changes in the volume fraction are observed between the top and bottom of the granular layer. As a consequence of the non-uniform vertical concentration profile, the upper regions tend to behave as a looser material than the layers located below. This behavior is accentuated  for the loose case with $E=0.1Pa$ (see figures \ref{fig:Sensi_E} and  \ref{fig:Sensi_E2}) where the difference in the initial concentration between the top and bottom granular layer is around $\Delta  \phi= \phi_{bottom} - \phi_{top} \approx 0.04$. Therefore, it is unsurprising that  the pore pressure values registered at the bottom for  the loose case using  $E=0.1Pa$  are shifted downwards (only negative values are predicted)  because that region is denser than the one corresponding to the $E=100Pa$ case.

\subsection{Influence of permeability}\label{permSensiSensiAnnex}

\begin{figure}
\centering
\includegraphics[width=0.95\textwidth]{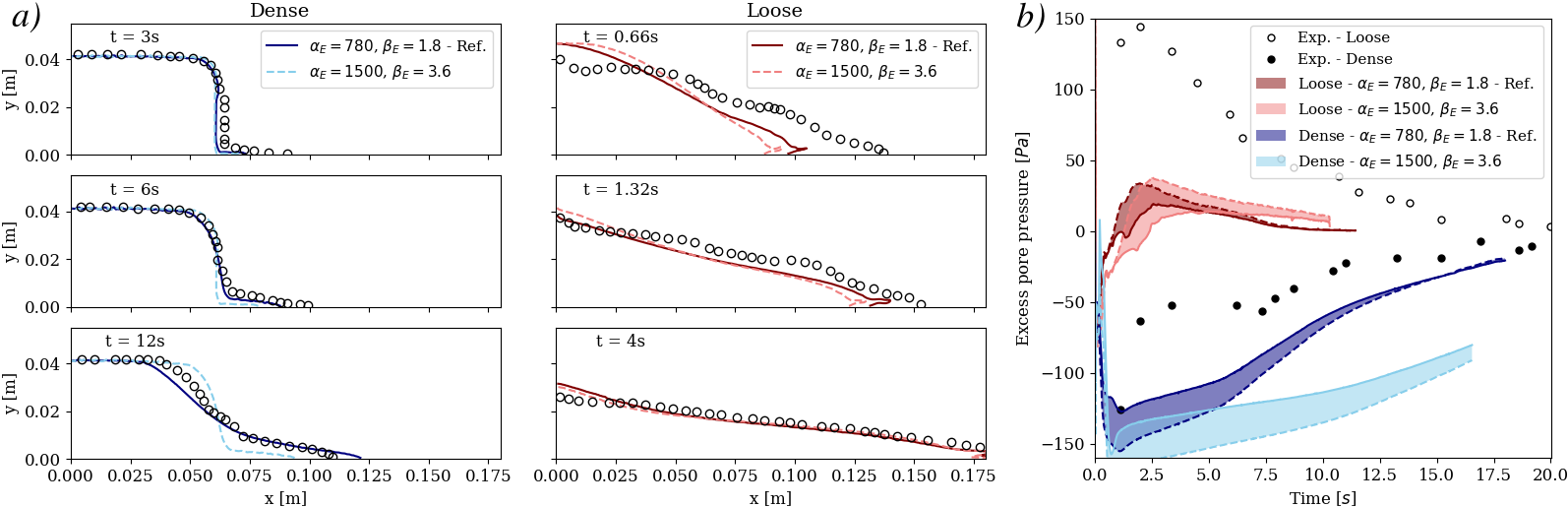}
\caption{\label{fig:Sensi_perm}Evolution of the morphology during the collapse of an initially dense and loose columns. b) Evolution of basal pore pressure measured at $2cm$ ($-$ continuous line) and  $3cm$ ($--$ dashed line). Shaded areas correspond to the region between the two probes results. The influence of Engelund's coefficients is investigated.}
\end{figure}

In order to study the role of permeability, Engelund's coefficients in Eq. \ref{eq:dragEngelund} are increased to lower the permeability of the granular material. More precisely, $\alpha_E=1500$ and $\beta_E=780$ are considered. Figure \ref{fig:Sensi_perm} shows lower permeabilities lead to a slow mobilization. In turns, the pressure dissipation takes longer as expected. Furthermore, figure \ref{fig:Sensi_perm} suggests different Engelund's coefficients have a negligible impact on the results for the loose granular column regarding the morphology and the pore pressure curve.

\subsection{Rheological coefficients}\label{rheologySensiAnnex}

\begin{figure}
\centering
\includegraphics[width=0.95\textwidth]{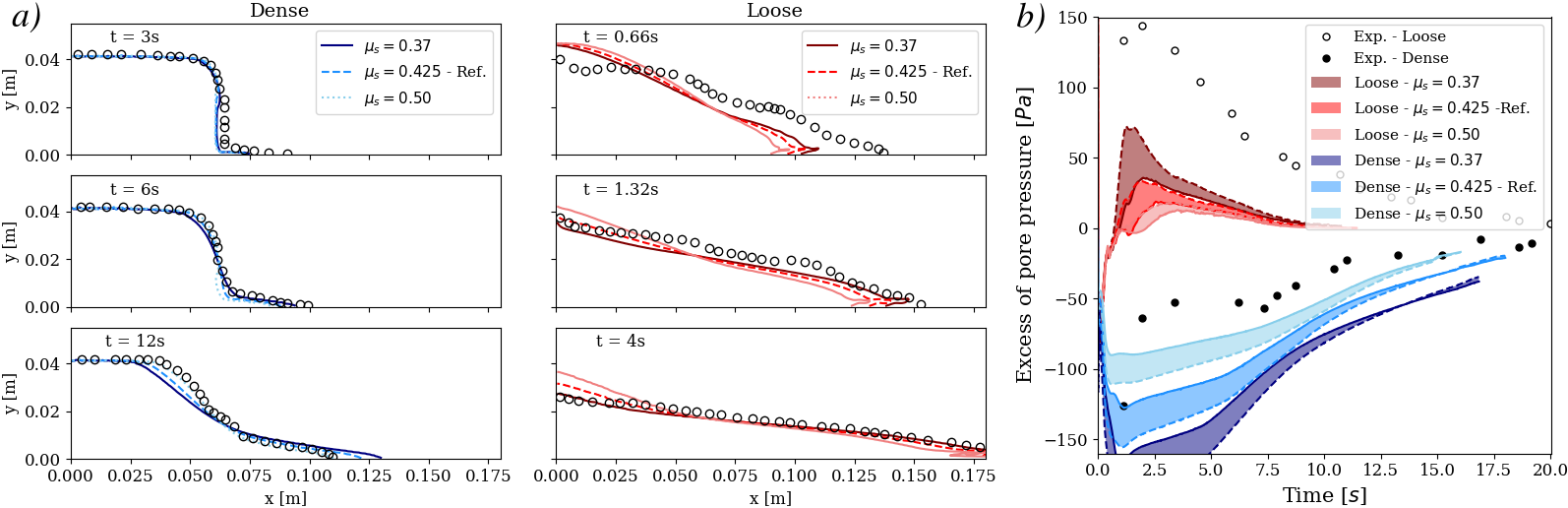}
\caption{\label{fig:Sensi_rheology}Evolution of the morphology during the collapse of an initially dense and loose columns. b) Evolution of basal pore pressure measured at $2cm$ ($-$ continuous line) and  $3cm$ ($--$ dashed line). Shaded areas correspond to the region between the two probes results.  Different rheological coefficients  are considered.}
\end{figure}

Figure \ref{fig:Sensi_rheology} shows large friction coefficients delay the collapse and,  at the end of the simulation, the deposit is characterized by a steeper slope. The lower mobilization entails a weaker pore pressure feedback: the soil is more difficult to shear, thus, pore volume changes take longer.  On the contrary, figure \ref{fig:Sensi_rheology} indicates low fiction angles enhance a rapid failure with abrupt pore volume changes, therefore, higher pore pressure jumps are observed for both initially dense and loose granular columns.

\section{Supplementary material: Mesh convergence test}\label{MeshAnnex}

\begin{figure}
\centering
\includegraphics[width=0.8\textwidth]{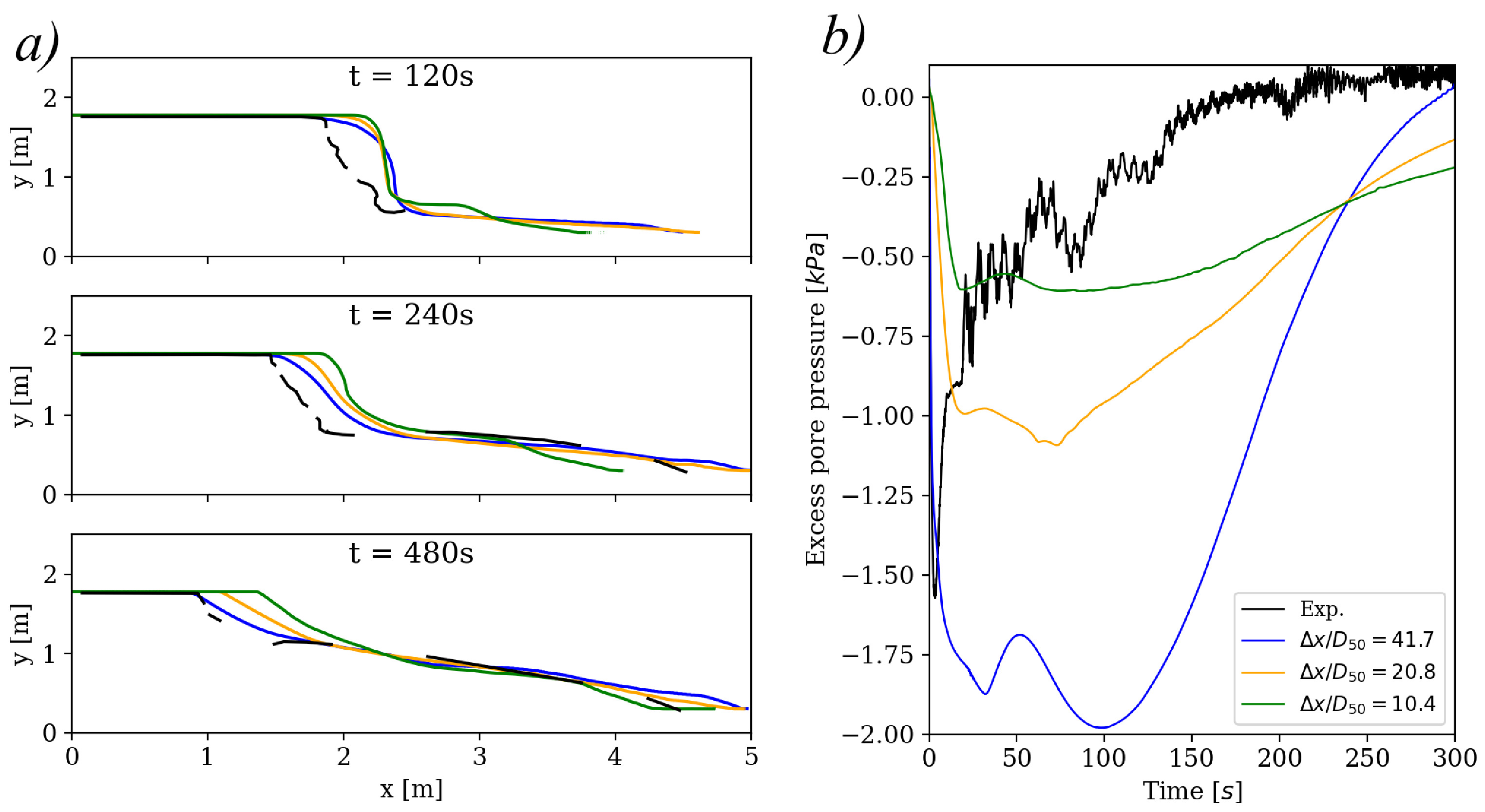}
\caption{\label{fig:MeshConvergence} a)  Comparison of the morphology and b)  pore pressure ($p^f$) evolution within the granular column  between the experiments and the numerical simulations for the GEBA sand. Three mesh sizes are considered.}
\end{figure}

Adopting different mesh sizes in the numerical model  has a strong influence on the accuracy of the results. In this analysis, three mesh sizes are considered. Figure \ref{fig:MeshConvergence}a shows the amount of rounding observed at the top of the breach wall is highly dependent on the grid size, with more rounding observed for larger grid cells and sharper corners for finer meshes. The time series displayed in figure \ref{fig:MeshConvergence}a also evidences an increment of the wall velocity as grid cells become larger. Finally, figure \ref{fig:MeshConvergence}b shows larger pressure jumps are predicted as the grid size increases. Moreover, pore pressure is dissipated more rapidly for coarser meshes.  The same pattern was reported by \cite{weij2020modelling}. The mesh refinement study displayed in figure \ref{fig:MeshConvergence}  is not converged, however,  finer meshes render the computation unfeasible for realistic simulations. In spite of the evident discrepancies in the pore pressure curves observed in figure  \ref{fig:MeshConvergence}b, the shape of the deposit and the wall velocity show reasonable differences between the $\Delta x /D_{50}=20.8$ and the $\Delta x /D_{50}=10.4$  grid sizes.

\end{document}